\documentclass{article}

\usepackage{complex-systems}
\usepackage{epsfig}

\usepackage{algorithm}
\usepackage{algorithmic}

\usepackage{fancyvrb} 

\usepackage{amsmath}
\usepackage{amsfonts}
\usepackage{amssymb}

\newcommand{\FP}{\mbox{ .}}     		
\newcommand{\COMMA}{\mbox{ ,}}     		

\newcommand{\R}{\mathbb{R}} 			
\newcommand{\Z}{\mathbb{Z}} 			
\newcommand{\N}{\mathbb{N}} 			

\newcommand{\U}{{\cal L}} 			

\newcommand{\fleche}{\rightarrow}		
\newcommand{\PartsOf}[1]{{\cal P}({#1})}	
\newcommand{\eq}{\thicksim}			
\newcommand{\function}[3]{ {#1} : {#2} \fleche  {#3} } 	
\newcommand{\set}[1]{\{#1\}}

\newcommand{\dens}{ {\rho} }

\newcommand{\x}[1]{{x(#1)}}
\newcommand{\y}[1]{{y(#1)}}

\newcommand{\dini}{d_{ini}}
\newcommand{\sr}{\alpha}

\newcommand{\ra}{ {r_a} } 
\newcommand{\rb}{ {r_b} } 
\newcommand{\df}{ {d_{avr}} } 
\newcommand{\muf}{ {\mu_{exp}} } 

\newcommand{\dyn}{{\Delta}}

\newcommand{\ECA}[1]{\textbf{#1}}

\newcommand{\Tt}{ T_{transient} }
\newcommand{\Ts}{ T_{sampling} }

\newcommand{\Word}[1]{\texttt{#1}}
\newcommand{\stzero}{\Word{0}}
\newcommand{\stone}{\Word{1}}
\newcommand{\zero}{ \bar{\stzero} }
\newcommand{\one}{ \bar{\stone} }

\newcommand{\Num}[1]{ \#_1( {#1} ) }

\newcommand{\D}{ n }

\newcommand{\E}{ Q ^{\U } }

\newcommand{\DebutTableau}[1]{
\begin{center}
\begin{tabular}{#1}
\hline
}

\newcommand{\FinTableau}{
\hline
\end{tabular}
\end{center}
}

\newcommand{\SzA}{8}

\usepackage{subfigure}
\newcommand{\BF}{\begin{figure}[htbp] \begin{center} }
\newcommand{\EF}{\end{center} \end{figure} }

\newcommand{\InsertFig}[2]{
	\centering	\epsfig{file=#1.eps, width=#2 cm}
}

\newcommand{\InsertTriFig}[4]{	
 		\InsertFig{#1}{#4}
 		\InsertFig{#2}{#4}
		\InsertFig{#3}{#4}			
}

\newcommand{\CaptionLabel}[2]{
	\caption{#1} \label{Fi:#2}	
}

\newcommand{\DoubleCaptionLabel}[3]{
	\caption{{\bf (a)} #1  {\bf (b)} #2 } \label{Fi:#3}	
}

\newcommand{\SSurf}[1]{
	\subfigure[ ]{ 
	\centering	\InsertFig{#1}{\SzA} 
	}
}

\newcommand{\DoubleOrbit}[2]{
	\subfigure[ ]{ 
		\epsfig{file=#1.eps, width=4 cm, clip=, bbllx=3.2cm, bblly=1.0cm, bburx=10.5 cm, bbury= 8cm}
		\epsfig{file=#2.eps, width=4 cm, clip=, bbllx=3.2cm, bblly=1.0cm, bburx=10.5 cm, bbury= 8cm}	 	
	}
}

\newcommand{\TripleOrbit}[3]{
	\subfigure[ ]{ 
		\epsfig{file=#1.eps, width=3.5 cm, clip=, bbllx=3.2cm, bblly=1.0cm, bburx=10.5 cm, bbury= 8cm}
		\epsfig{file=#2.eps, width=3.5 cm, clip=, bbllx=3.2cm, bblly=1.0cm, bburx=10.5 cm, bbury= 8cm}
		\epsfig{file=#3.eps, width=3.5 cm, clip=, bbllx=3.2cm, bblly=1.0cm, bburx=10.5 cm, bbury= 8cm}
	}
}

\newcommand{\LargeOrbit}[1]{
		\epsfig{file=#1.eps, width=3.2 cm, clip=, bbllx=2.6cm, bblly=1.0cm, bburx=9.5 cm, bbury= 14cm}
}

\newcommand{\figref}[1]{Figure~\ref{Fi:#1}}

\newcommand{\BeginList}{\begin{itemize}}
\newcommand{\EndList}{\end{itemize}}
\newcommand{\BL}{\BeginList}
\newcommand{\EL}{\EndList}

\newcommand{\BQ}{\begin{quote}}
\newcommand{\EQ}{\end{quote}}

\newcommand{\LegSurfBifZNa}{An example of surface with a discontinuity at $ \sr = 1.0 $ and noise for $ \sr < 1.0 $~: ECA \ECA{73}
(z-axis inverted for allowing the display of discontinuity at $\sr$ = 1).}
\newcommand{\LegSurfBifZNb}{Evolution of ECA \ECA{73}~: (left) $ \sr= 1.0 $ (right) $ \sr = 0.8 $.}

\newcommand{\LegSurfHorizontala}{An example of horizontal surface~: ECA \ECA{90}}
\newcommand{\LegSurfHorizontalb}{Evolution of ECA \ECA{90}~: (left) $ \sr= 1.0 $ (right) $ \sr = 0.5 $. In this space-time diagram and in the following the intial condition is obtained with a Bernoulli process with $ \dini = 0.5 $, the grid size is $ \D = 50 $, the time is from $ t = 0 $ to $ t= 49 $.}

\newcommand{\LegSurfBIFZa}{An example of GAP model sampling surface (z-axis rescaled)~: ECA \ECA{2}.}
\newcommand{\LegSurfBIFZb}{Evolution of ECA \ECA{2}~: (left) $ \sr= 1.0 $ (right) $ \sr = 0.5 $.}

\newcommand{\LegSurfMaja}{An example of $ \dini $-dependent, $ \sr $-invariant sampling surface~: ECA \ECA{232}.}
\newcommand{\LegSurfMajb}{Evolution of ECA \ECA{232}~: (left) $ \sr= 1.0 $ (right) $ \sr = 0.5 $.
A tight examination of the configuration shows that the width of the second white band is larger in the left diagram.}

\newcommand{\LegSurfBIFOa}{Sampling surface for an SPT model~: \ECA{50} (z-axis rescaled). }
\newcommand{\LegSurfBIFOb}{Evolution of ECA \ECA{50}~: (left) $ \sr= 1.0 $ (center) $ \sr = 0.75 $ (right) $ \sr = 0.25 $. }

\newcommand{\LegSurfContia}{An example of $ \dini $-invariant,$ \sr $-dependent sampling surface~: ECA \ECA{126} ($\sr$-axis inverted).}
\newcommand{\LegSurfContib}{Evolution of ECA \ECA{126}~: (left) $ \sr= 1.0 $ (center) $ \sr = 0.9 $ (right) $ \sr = 0.5 $.}

\newcommand{\LegSurfShifta}{An example of ill-defined surface~: ECA \ECA{170} (shift).}
\newcommand{\LegSurfShiftb}{Evolution of ECA \ECA{170}~: (left) $ \sr = 1.0 $ (right) $ \sr = 0.8 $.}

\newcommand{\LegSurfUa}{An example of a U-shaped sampling surface~: ECA \ECA{6}. }
\newcommand{\LegSurfUb}{Evolution of ECA \ECA{6}~: (left) $ \sr= 1.0 $ (center) $ \sr = 0.75 $ (right) $ \sr = 0.25 $. }

\newcommand{\LegSurfRiddlesa}{An example of sampling surface with ``riddles''~: \ECA{46} ($\sr$-axis inverted). }
\newcommand{\LegSurfRiddlesb}{Evolution of ECA \ECA{46}~: (left) $ \sr= 1.0 $ (center) $ \sr = 0.75 $ (right) $ \sr = 0.25 $. }

\begin{document}

\title{An Experimental Study of Robustness to Asynchronism for 
\\Elementary Cellular Automata}

\author{\authname{Nazim A. Fat{\`e}s%
\thanks{
Electronic mail address:  \texttt{Nazim.Fates@ens-lyon.fr} 
}}
\\[2pt] 
\authadd{Laboratoire de l'Informatique du Parall\'elisme, ENS Lyon, 46, all\'ee d'Italie}\\ 
\authadd{ 69 364 Lyon Cedex 07 - France }%
\\[2pt]
\and
\authname{Michel Morvan
\thanks{
Electronic mail address:  \texttt{Michel.Morvan@ens-lyon.fr}
}
} \\ 
[2pt]
\authadd{Laboratoire de l'Informatique du Parall\'elisme, ENS Lyon, 46, all\'ee d'Italie}\\ 
\authadd{ 69 364 Lyon Cedex 07 - France }%
}

\maketitle

\begin{abstract}
Cellular Automata (CA) are a class of discrete dynamical systems that have been widely used to model complex systems 
in which the dynamics is specified at local cell-scale. 
Classically, CA are run on a regular lattice and with perfect synchronicity. 
However, these two assumptions have little 
chance to truthfully represent what happens at the microscopic scale for physical, biological or social systems.
One may thus wonder whether CA do keep their behavior when submitted to small perturbations of synchronicity.

This work focuses on the study of one-dimensional (1D) asynchronous CA with two states and nearest-neighbors. 
We define what we mean by ``the behavior of CA is robust to asynchronism'' using a statistical approach with macroscopic parameters
and we present an experimental protocol aimed at finding which are the robust 1D elementary CA.  
To conclude, we examine how the results exposed can be used as a guideline for the research of suitable models according to robustness criteria.
\end{abstract}

\section{Introduction}

The aim of this article is to study the robustness to asynchronism for cellular automata. 
In other words, we propose to examine some qualitative and quantitative aspects of the change of behavior that are induced 
when the cells are no longer updating their state systematically at each time step.

The first study of the effect of asynchronism was carried out in 1984 by Ingerson and Buvel in \cite{Ingerson84}~:
\begin{quote}
`` (...) Cellular automata exhibit such remarkable self-organization that it is certainly tempting to consider the possibility that they may be a valid model for real-world systems, such as the growth of biological organisms, crystals, snowflakes, etc. However, one commonly made assumption about these systems is that the cell iterate synchronously. We wanted to estimate how much of the interesting behavior of cellular automata comes from synchronous modeling and how much is intrinsic to the iteration process.'' 
\end{quote}
The authors carried out experiments on the space of ``elementary cellular automata'' rules (see \ref{Sec:ECA}) 
and showed that varying the iteration process produced significant change in the evolution of some cellular automata
whereas some other cellular automata were not affected by the modifications. 
The study was however purely qualitative and no algorithmic method was proposed to systematically estimate these changes.

In 1993, Huberman and Glance criticized the use of CA as a modeling tool 
that could be suitable for describing real-world phenomena \cite{Huberman93}.
The model they studied is a spatially-extended version of the prisoner's dilemma ``with no memories among players and no strategical elaboration'' 
introduced by Nowak and May in \cite{Nowak92}.
They argued that the model was not realistic because it used the assumption that the actors all updated their strategy synchronously. 
Their experiments showed that when the perfect synchrony assumption was dropped, significant changes of behavior were observed.
At the same time, similar ideas were developed by Stark in the field of biology\cite{Sta94}.

In 1994, Bersini and Detours studied an asynchronous version of the Game of Life \cite{Ber94}. 
They observed that the introduction of asynchrony led to modifying the dynamics from a behavior with long transients 
to a behavior with fixed points. The authors explained this property by identifying some asynchronous CA with Hopfield neural network and by
proposing a description of the asynchronous behavior in terms of Lyapunov energy functions. 
This raised the question to know whether the stabilization effects was to be observed for any model or was specific to the models chosen by the authors. 
This article partially answers this question by exhibiting counter-examples for which the increase of asynchronism 
leads to less stability (see Section \ref{Sec:Bersini}).

The first quantitative study of the influence of the way transitions were made in CA 
were carried out by Sch\"onfisch and de Roos \cite{Sch99}.
The authors use explicit functions for updating the cells and 
show that the evolution of a cellular automaton might strongly
depend on the correlation between the spatial arrangement of cells and the order of their update.
For example, if the cells are arranged in a line, one could consider the possibility of updating the cells one-by-one from left to right.
The correlation between the updating method order and the spatial position of the cells is analytically estimated and it appears
that for some type of updating methods, the evolution of the cellular automaton becomes strongly dependent on the lattice size.
The important result is that among the different update methods studied, 
the only method which did not introduce any spurious correlations 
consisted in choosing the cells of the lattice randomly with an equal probability for each cell.
In the study we here present, we only consider this particular type of asynchronism and rather concentrate on
the study of the phenomenological changes observed.

The purpose of this work is to propose a first algorithmic approach to answer to the question 
``To which extent is the behavior of cellular automaton dependent on the synchrony of the transitions?''. 
In other words, we  want to know if the application of a small change in the way the transitions are performed 
leads to brutal changes of the ``behavior''. 
Note that this differs from studying the effect of perturbing the configuration themselves, for example introducing noise in the system.
In Section \ref{Sec:Definitions}, we give formal definitions of the CA concepts and we describe the algorithm we use to quantify CA robustness.
In Section \ref{Sec:ECAStudy}, we analyze the results by sorting the models according to the robustness quantification given by our protocol. 
In the last section, we discuss the results and analyze how the study of robustness could be related to the activity of 
modeling complex systems with CA.
\section{ Definitions and experimental protocol } \label{Sec:Definitions}

In this section, we formally define the notion of asynchronous cellular automaton. 
We then describe the experimental protocol used to quantify the robustness of a model using the notion of ``sampling surface'' 
and the notion of ``robustness indicator''.
Finally, we analyze some intrinsic limits of our protocol.

\subsection{ Asynchronous Cellular Automata } \label{Sec:definitions}

An {\em Asynchronous Cellular Automaton} (ACA) is a 5-tuple $ (\U, Q, G, f, \dyn) $ defined as follows~:

\BeginList
\item	A {\em cell} is a variable that takes its values in  $ Q $, the set of possible {\em states}. 

\item	The set of all cells is called the {\em lattice}, it is denoted by $ \U $ and we have $ \U \subseteq \Z^d   $, 
where $ d $ is the {\em dimension} of the lattice.

\item	The {\em neighborhood} of a cell $ N(c) $ is a function which associates to a cell $ c $ an ordered set of cells. 
The cardinality of $ N(c) $ is constant and is equal to  $ N $.

\item	$ \function{f}{Q^{N}}{Q} $ is {\em the local transition rule} which defines 
	how a cell updates its state according to the states of the cells located in its neighborhood.

\item 	$  \function{\dyn}{\N}{\PartsOf{\U}} $ is the {\em updating method} \cite{Sch99}, which defines for each time $ t $, 
	the set of cells to which the transition rule is applied.
	In a modeling approach, $ \dyn $ might be seen as defining the set of non-defective cells at time $ t $, 
	with the convention that a defective cell will keep its state constant
	whereas a non-defective cell will update its state according to the local rule. 
	
\EndList

The updating method $ \dyn $ is said to be {\em synchronous} 
if $ \forall t, \dyn(t) = \U $,  otherwise it is {\em asynchronous}.	
In this context, it appears that ``classical'' Cellular Automata form a particular sub-class of ACA, 
for which the update rule is synchronous. 
We restrict here our study of updating methods to the sub-class of {\em step-driven methods} \cite{Sch99}, 
in which the expression of time does not appear explicitly in the definition of $ \dyn $. 
Among all the possible step-driven methods, 
we choose to use  {\em asynchronous stochastic dynamics}, denoted by $ \dyn_{\sr} $,
defined by considering for each time $ t $  every cell of $ \U $ and assigning a probability $ \sr $ that this cell is in $ \dyn(t) $. 
The parameter $ \sr \in ]0,1]$  is called the {\em synchrony rate}.
This updating method has the advantage of satisfying a ``fair sampling condition''
which specifies that each cell should be updated an infinite number of times without any bias\footnote{ 
In \cite{Bahi02}, the definition of the ``fair sampling condition'' 
only imposes that each cell should be updated an infinite number of times.
In our context, we have chosen to add the property that each cell should also be chosen with an equal probability to any other cell.
} ~: \label{Sec:FairSampling}
\[ \forall c \in \U, \lim_{T \fleche \infty}{ \frac{ card \mbox{ } \set{ t \leq T, c \in \dyn(t) } }{T} } 
= \frac{\sr}{card \mbox{ }\U} \FP \]

An assignment of a state to each cell of $ \U $ is called a {\em configuration}. 
It is denoted by $ x = { (\x{c})_{ c \in \U } } $, with $ x \in \E $.
$ \dyn $ being fixed, the {\em global transition function} is a function  $ \function{F_{\dyn}}{\E \times \N}{\E} $ 
which associates to each 
configuration $ x = { ( \x{c} )_{ c \in \U } } $ and to each time $ t$,
a configuration $ y = { ( \y{c} )_{ c \in \U }  } $
such that~:
\BeginList
	\item $ \y{c} = f[ N(c) ] $ if $ c \in \dyn(t) $
	\item $ \y{c} = \x{c} $ otherwise.
\EndList

\begin{figure}[hpbt] 
	\begin{center}
	\centering	 
	\epsfig{file=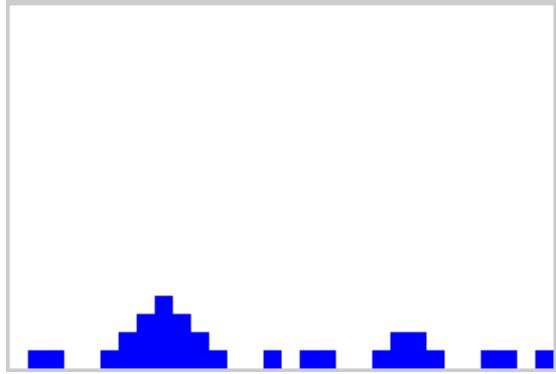, width=\SzA cm, clip=, bbllx=0.6cm, bblly=0.5cm, bburx=9 cm, bbury= 8cm}
	\caption{Example of space-time diagram of ECA \ECA{128} (see \ref{Sec:ECA} for coding). 
Configurations are displayed horizontally and time goes from bottom to top. } \label{Fi:SPTfig}
	\end{center} 
\end{figure}

A global transition function is a particular kind of discrete dynamical system acting on configurations.
We thus can associate to each configuration $ x $ its {\em orbit}, 
the series of configurations $ (\gamma_{\sr}(x,t))_{t \in \N} $ obtained by the iteration of $ F_{\dyn_{\sr}} $ on $ x $ 
using the recursive definition $ \gamma_{\sr}(x,t+1) = F_{\dyn} ( \gamma_{\sr}(x,t), t) $. 
However, unlike dynamical systems which do not depend on an update function, 
when the updating method is not synchronous (i.e, when $ \sr < 1 $),
the orbit of $ \gamma_{\sr}(x,1) $ is {\em not} necessarily the shifted orbit of $ x= \gamma_{\sr}(x,0) $. 
We will say that a configuration $ x_f $ is a {\em fixed point} if $ \forall \dyn, \forall t, F_{\dyn} (x, t) = x $.
Finite parts of the orbits can be represented in space-time diagrams, where configurations are represented horizontally and
where time is represented vertically (see \figref{SPTfig}).

In the sequel, we will be interested in some configurations in which the transition of information is blocked. 
We say that a word $ w \in Q^*  $ is a {\em wall} if it verifies : $ \forall (u,v) \in Q \times Q, F_{|w}[uwv] = w $, 
where $ F_{|w} $ denotes the restriction of $ F $ on the cells that compose $ w $. 
A wall is a ``strong'' type of {\em blocking word} (i.e., a word that splits a configuration 
into two parts by preventing any information to cross it \cite{Var03}). 
We will call a {\em q-domain} a set of adjacent cells that are all in state $ q $.

\subsection{One dimensional Elementary Cellular Automata} \label{Sec:ECA}

In this paper, we restrict our study to the one-dimensional case, taking $ d = 1 $. 
We call {\em elementary cellular automata} the class of 1D-ACA defined by  
$ Q = \set{ \stzero, \stone } $ and $ \forall c \in \Z, N(c)= \set{ c-1, c, c+1 } $. 
As the study is experimental, we only consider lattices of finite size, using  {\em periodic boundary conditions}~: 
1D lattices are {\em rings} and indices of $ \U $ are taken in $ \Z / { \D \Z } $, with $ \D $ size of the ring.

Following \cite{Wol83} we associate to each ECA $ f $ its code~:
\[ W(f) = f(0,0,0) \cdot 2^0 + f(0,0,1) \cdot 2^1 + \cdots + f(1,1,0) \cdot 2^6 + f(1,1,1) \cdot 2^7 \FP \]
An ECA $ f $ having the code $ R = W(f) $  is denoted by ECA \ECA{R} and we will equally use the more general word 'model' to qualify a rule.
The symmetry operations obtained by the left/right exchanging and $\stzero$/$\stone$ complementation allow to associate to each rule \ECA{R},
a reflected rule \ECA{Rp}, a conjugate rule  \ECA{Rc}, and a reflexive-conjugate rule  \ECA{Rcp}. 
The association of \ECA{R} to \ECA{(R, Rc, Rp, Rcp)} allows the partition of the ECA space into 88 equivalence classes
and we will call {\em minimal representative} the rule that has the smallest index in a class.
In the sequel, we will work in this quotiented space and we will only consider minimal representative rules.

\subsection{ Experimental protocol for robustness estimation } \label{Sec:protocol}

The purpose of this section is to introduce formal notations that allow to specify 
the protocol we use to obtain the experimental data.
We then introduce the concept of 'sampling surface' to qualitatively estimate a model's robustness to asynchronism 
and we propose to quantify this robustness using two parameters. 
Finally, we analyze the limits of our protocol.

\subsubsection{Definition of the protocol}

The macroscopic measures we use to estimate the change of behavior of an ECA 
are based on the statistical analysis of the density variations.
The {\em density} of a configuration is a real number defined by $\function{\rho}{\E}{[0,1]}$ 
such that $ \rho(x) = \frac { \Num{x} }{ |x| } $ 
where $ \Num{x} $ denotes the number of $ \stone $'s in $ x $ and $ |x| $ is the size of the configuration $ x $.
In a previous work \cite{Fates03}, we showed that the density and more precisely the evolution of the density can be considered 
as a pertinent parameter for describing in a first approximation the global behavior of an ECA. 
For example, it can be used as a means 
of discriminating the chaotic-looking ECA from the regular-looking ones.

The density is used here to identify the models that are non-robust to the introduction of asynchronism.
In order to have an ``observation function'' $ \mu $ that will quantify changes in behavior, 
we use an experimental protocol that depends on five parameters~:

\BL
\item The size of the grid $ \D $.
\item The density of the initial condition $ \dini $.
The initial condition $ x(\dini) $ is constructed using a Bernoulli process~: 
for every cell of $ x $, this cell has a probability $ \dini $ to have state $ \stone $ and a probability $ 1 - \dini $ to have state $ \stzero $. 
The distribution of the density of $ x $ is binomial, 
which implies that $ d(x) $ is close to $ \dini $ for large $ |x| $ with high probability but note that it is not often strictly equal to $ \dini $. 
\item The synchrony rate of the update method $ \sr $.
\item The transient time $ \Tt $ after which the orbits are analyzed.
\item The sampling time $ \Ts $ during which the orbits are analyzed.
\EL

In order to obtain $ \mu $ experimentally, we take the initial condition $ x(\dini) $
and let the ACA defined with synchrony rate $ \sr $ evolve during $ \Tt $ steps.
We then store the value of the density during $ \Ts $ steps and average this value to obtain $ \muf ( \dini, \sr ) $~:

\[ \muf( \dini, \sr)  = 
  \frac{1}{\Ts} \sum_{t=\Tt + 1}^{t= \Tt + \Ts} d( \gamma_{\sr}( x(\dini), t) ) \FP \]

\begin{algorithm}
\caption{Construction of a sampling surface} \label{algoSS}
\begin{algorithmic}

\FOR{ $ d_{val} $  = $ d_{min} $ to $ d_{max} $ step $ d_{stp} $ }
\STATE $ x_{ini}(d_{val}) \leftarrow $ random initial condition of density ($ d_{val} $)
\ENDFOR

\FOR{ $ \sr $ = $ \sr_{min} $ to $ \sr_{max}$   step $ \sr_{stp} $  }
\FOR{ $ \dini $ =  $ d_{min} $ to $ d_{max} $  step $ d_{stp} $  }

\STATE $ 	x \leftarrow x_{ini}( \dini ) $ // initial condition 
\FOR{ $ t_1$  = 1 to $ \Tt $ }
\STATE		$ x \leftarrow F_{\sr}(x, t_1) $
\ENDFOR
\FOR{ $ t_2 $ = 1 to $ \Ts $ }
\STATE	     $	x \leftarrow F_{\sr}(x, t_2) $
\STATE	     $ \mbox{ sample[ t2 ] } \leftarrow \dens( x ) $
\ENDFOR
\STATE	$ \df(\sr, \dini) \leftarrow \mbox{ Average[ sample ] }$

\ENDFOR
\ENDFOR

\end{algorithmic}
\end{algorithm}

Exhaustive experimentation on all initial conditions and all values of synchrony rate are impossible in practice. 
This means that we have to do a sampling by  randomly choosing some initial conditions and some synchrony rates.
For the initial densities, we choose to perform a {\em uniform density} sampling :\\ 
We construct our set of initial densities
$ D $ with values varying from $ d_{min} $ to $ d_{max} $ with step $ d_{stp} $. 
We denote this kind of interval by $ D = [ d_{min}, d_{max}] (d_{stp}) $.
Similarly, we construct our set of synchrony rates by taking
$ A = [ \sr_{min}, \sr_{max}] (\sr_{stp}) $, 
with $ \sr_{max}= 1.0 $ (the synchronous case is sampled).

The sampling operation thus results in the application of {\em Algorithm~\ref{algoSS}} and 
its output is a set of points $ \muf(\dini, \sr )$ with $ \dini \in D $ and $ \sr \in A $. 
It can be represented in a 3D space in the form of a two-dimensional {\em sampling surface}.

\BF
\SSurf{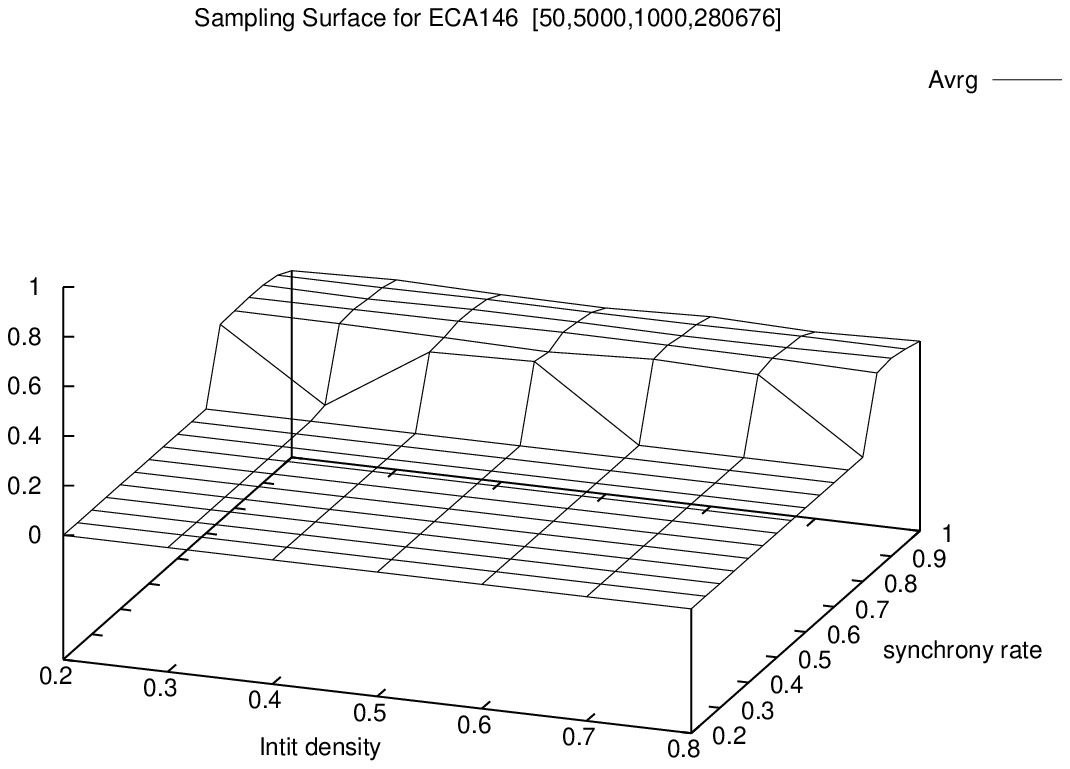}
\caption{An example of sampling surface : ECA \ECA{18}. The value of the indicators for this surface are $ \ra = 0.03 $ (no change around $ \sr \eq 1$) and $ \rb = 0.19 $ (important change in the asynchronous domain).}
\label{ExampleSS}
\EF

In order to obtain a first level of classification, we extract quantitative information from our sampling surfaces by
computing out two parameters from the experimental data~:
\BL
\item 	The first parameter is used to measure how small introduction of synchrony affects the global behavior of the CA. 
	The {\em small-asynchrony-introduction} indicator $ \ra $ is given by~:
	\[ \ra = \set{ \frac{1}{|D|} \sum_{ d \in D} {[ \muf( d, \sr_{as} ) - \muf( d, 1.0) ]}^2 }^{1/2} \FP \]
	This parameter is the quadratic average of the variations of $ \muf $  
	between total synchronism  $ \sr_{as} = 1.0 - \sr_{stp} $ and the highest asynchronous value.
	It somehow estimates the averaged absolute value of the ``jump'' of value that can occur for $ \sr \eq 1 $.

\item The second parameter is used to measure how the change of synchrony from $ \sr_{as} $ to $ \sr_{min} $ 
globally affects the behavior of the CA.
The {\em asynchrony-dependence indicator} $ \rb $ is then defined by~:
\[ \rb = sup_{\sr  \in A' } \set{ \frac{1}{|D|} \sum_{ d \in D} {[ \muf( d, \sr + \sr_{stp} ) - \muf( d, \sr ) ]}^2 }^{1/2} \FP \] 
with $ A'= [ \sr_{min}, \sr_{max}-2.\sr_{stp}] (\sr_{stp})  $.
This parameter is the maximum quadratic average on $ \dini$, for all asynchronous densities, of the variations of $ \muf $.
It estimates, in the asynchronous regime, how far from invariance in respect to the translation of axis $ \sr $ the surface is.

\EL

\subsubsection{Limits of the protocol}

Like in any simulation approach, 
the width of validity for the results we obtain is limited by some of the choices we had to make in the design of the protocol. 
Let us try to identify some of these limits.

First, it is clear that our results are limited to the particular region defined by the constants chosen in the experimental protocol.
In order to calculate the ``observation function'', we have to set the value of the parameters $ \Tt$ and $ \Ts $.
These values are chosen as big as possible 
with the implicit assumption that $ \muf $ does no longer change when $ \Tt $ and $ \Ts $ are increased.

Similarly, the choice of the grid size  $ \D $ might influence the outcome of the results.
For example, the particular ECA \ECA{90} has a transition function that can be expressed in the synthetic form~: 
$ \forall (a,b,c) \in Q^3, f(a,b,c) = a \oplus c $ with $ \oplus $ denoting the addition modulo 2.
The additivity of the local rule allows a superposition principle to be obeyed by the global rule~:
\[ \forall (x,x') \in \E \times \E, F_{synch}( x \oplus x' ) = F_{synch}(x) \oplus F_{synch}(x') \FP \]
The evolution of configuration containing a single cell that is in state $ \stone $ 
leads to the formation of Pascal's Triangle modulo 2.
Using the superposition principle\label{Sec:superposition}, it is easy to see that for grid sizes that are powers of two, $ n = 2^k, k \in \N $, 
any initial configuration evolves to the null configuration $ \zero $ in number of step less or equal to  $ n/2 $. 
However, for sizes that are not powers of two this nilpotency property does not hold any more and we instead observe cycles 
whose length are only bounded by $ 2^n $ (see \cite{Martin84} for a more precise analysis).
This simple observation shows that we should be very careful not to generalize a result obtained on a particular ring size to any ring size.
We however conjecture that the experimental data are not dependent on the ring size for most of the ECA rules.
The experimental examination of this assumption will be done in the next section for a small number of values of $ \D $.

Let us also stress that the protocol associates to a given initial density {\em the same} initial configuration
which is re-used for different synchrony rates.
Moreover, we take only {\em one} sample for each couple of control parameters $ (\dini, \sr) $. 
Another possibility would consist in taking several samples for each point 
and then compute the average of the measured values $ \muf $.
However, this averaging effect could be misleading in the estimation of the model's robustness~: 
for some particular rules (e.g. shift) it would be possible to have a behavior that varies strongly according 
to the initial condition chosen but have a stable average. 
In this work, we choose to say that such a CA is {\em not} robust 
because we are interested in a concept of robustness that characterizes the evolution of a single configuration
and not subsets of configurations. This will be further discussed in Section \ref{Sec:Discussion}.

All these limitations clearly imply that the indicators $ (\ra, \rb) $ and
even the sampling surfaces are far from holding all the information about a model's behavior. 
They should instead be considered as a way of making a projection of the huge space of all possible orbits into the simpler $ \R ^ 2 $ space.
They can also be viewed as a first approximation tool to identify the ``non-robust'' CA. 
Indeed, if a perturbation produces a change in the density distribution then we are allowed to affirm 
that we are in presence of a change in behavior.  
The converse is not true since one could easily imagine a situation in which the
density distributions would stay stable whereas some other macroscopic parameters would vary. 
So there are at least two other limitations
of the protocol proposed~: the first one is that the use of the density induces a compression of information
that could introduce biases for behavior estimation, especially when a rule is number conserving (i.e., when its evolution conserves the density). 
The second one is that we rely on two indicators that are chosen as quantifiers of the regularity of the sampling surfaces 
using again an approximation. 
The analysis of experimental results is then a three-level analysis~: 
the first and second one are qualitative, they consist in the visual examination of 
the space-time diagrams and the sampling surface. 
The third one is quantitative and uses the indicators $ (\ra, \rb) $. 
These restrictions confirm once more that this work is just a first step in the study of asynchronous robustness. 
It aims to give a global view of the landscape in order to show the pertinence of the problem and to identify to some challenging ways to explore.

\section{Exhaustive study of the ECA space } \label{Sec:ECAStudy}

In this section, we start by examining the repartition of all ECA into the indicators space, and divide this space into zones.
For each zone, we show the sampling surfaces and 
we examine how the dynamical systems actually evolve by looking at some orbits.

\subsection{Repartition of the ECA}

The results were obtained with the experimental value for 
transient time $ \Tt = 5000 $, 
sampling time $ \Ts = 1000 $, 
ring size  $ \D = 50 $, 
initial density sampling interval $ D=[0.2, 0.8](0.1) $, 
synchrony rate sampling interval $ A=[0.2, 1.0] (0.1) $.
All the experimental data were obtained with a software dedicated to the study of CA robustness \cite{FiatLux}.

\BF
\InsertTriFig{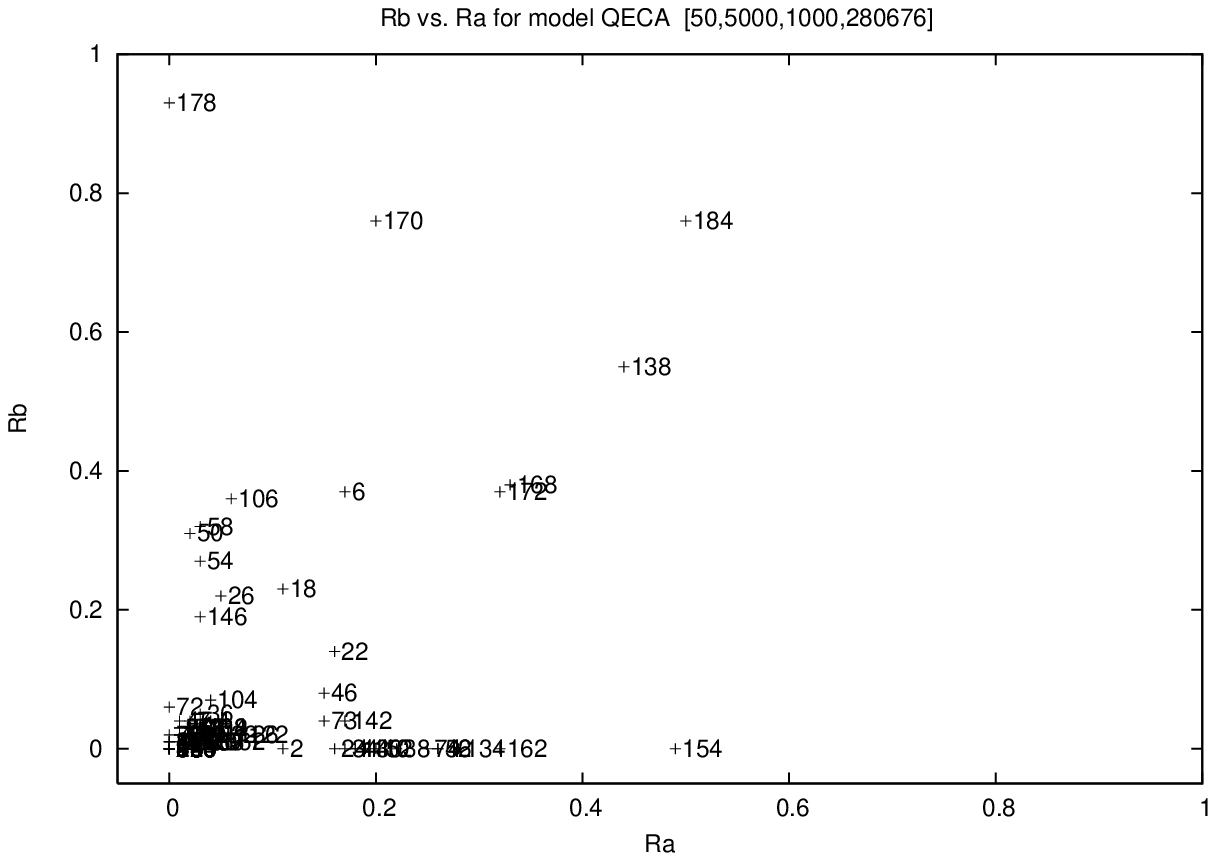}{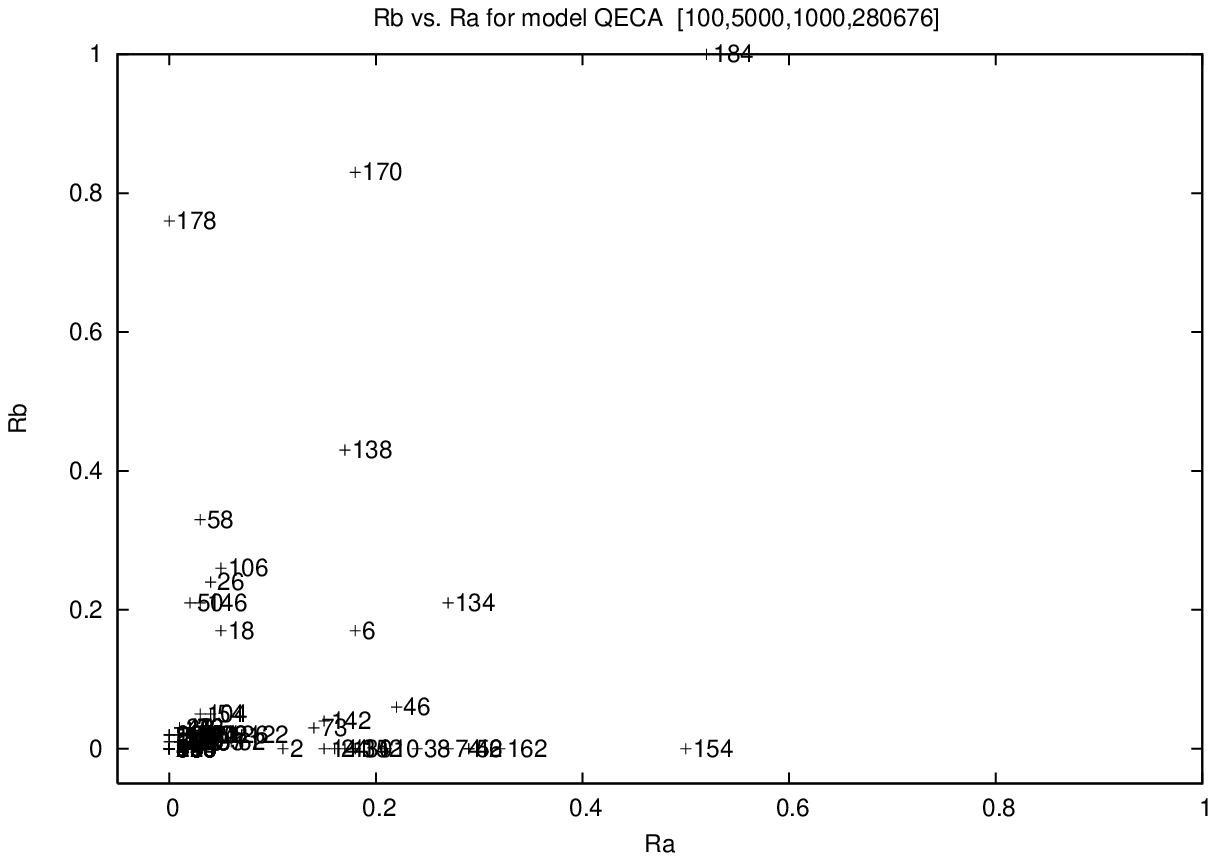}{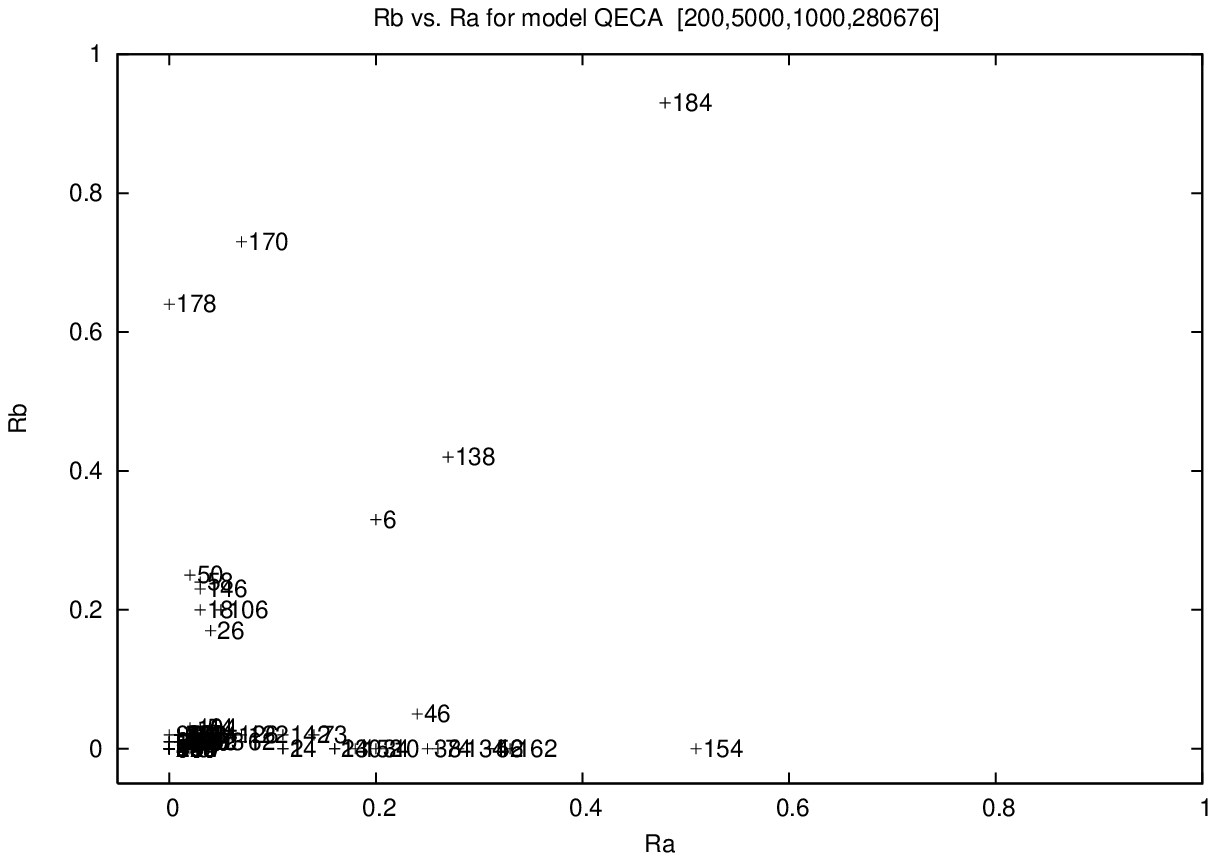}{\SzA}
\CaptionLabel{Evolution of the repartition in the space $ (\ra, \rb) $ according to different ring sizes~:
 $ \D = 50 $ (up), $ \D = 100 $  (middle), $ \D = 200 $  (bottom); transient and sampling times were~: 
$\Tt = 5000 $, $ \Ts = 1000 $. }{EvolRep}
\EF

\figref{EvolRep} shows the repartition\footnote{
Recall that only minimal representative ECA have a corresponding point in this space.
} of the ECA in the 2D space $ (\ra, \rb) $ for three different values of ring size $ \D $.
In the three diagrams, the dispersion of the ECA is far from uniform and rather forms groups. 
If the relative position of the points may vary from one value of $ \D $ to another, the diagrams appear to have similar distributions.
These observation leads us to consider in a first step that the diagram can partioned into 4 zones~:
\BeginList
\item In Zone A, we group the ECA that form a dense group in the region defined by $ \ra < 0.1 $ and $ \rb < 0.1 $. 
\item In Zone B, we group the ECA that stretch along the $ \ra $-axis : $ \ra > 0.1 $, big $ \rb < 0.1 $.
\item In Zone C, we group the ECA that stretch along the $ \rb $-axis : $ \ra < 0.1 $, big $ \rb > 0.1 $.
\item In Zone D, we group the other ECA : $ \ra > 0.1 $, $ \rb > 0.1 $.
\EndList

Let us now study each zone separately in order to see if the discrimination introduced by the $ \ra $ and $ \rb $ parameters
does allow to separate the ECA into meaningful classes.
For each zone, we examine the shape of the sampling surfaces obtained and try to analyze 
how this shape is related to the configurations found in the model's orbits.

\subsection{Zone A (small $ \ra $ and small $ \rb $)}

This zone contains the ECA with high robustness to asynchronism. 
The models in this zone are situated close to the point  $ (\ra, \rb) = (0, 0) $, 
this means that given a specific initial condition, 
the orbits obtained with different synchrony rate produced the same values for the observation function $ \muf $.
There can be two straightforward ways to explain this property~: 
\BL
\item (H1) The configurations of the asymptotic part (i.e., after the transient time is elapsed) of the orbits are different 
but the averaging effects used in the experimental protocol
produce identical measures for the observation function (see \ref{Sec:protocol}).
\item (H2) The configurations of the asymptotic part of the orbit are similar despite having different trajectories during the transient time.
\EL

\subsubsection{Horizontal surfaces}

\BF
\SSurf{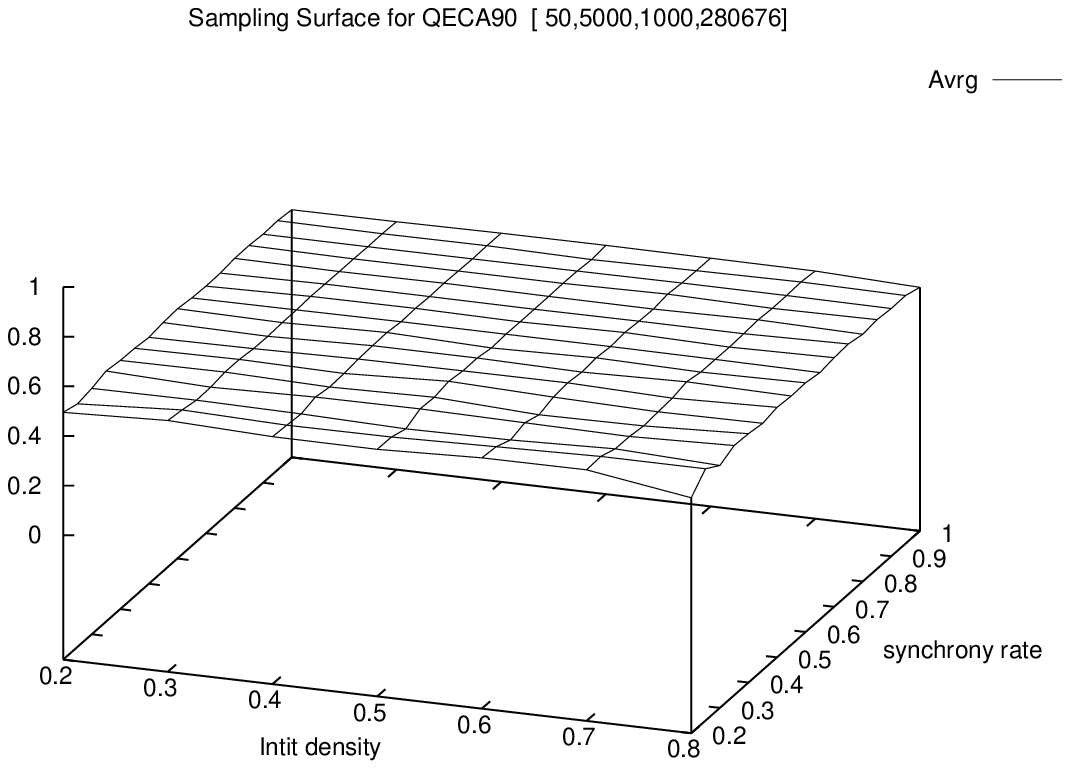}
\DoubleOrbit{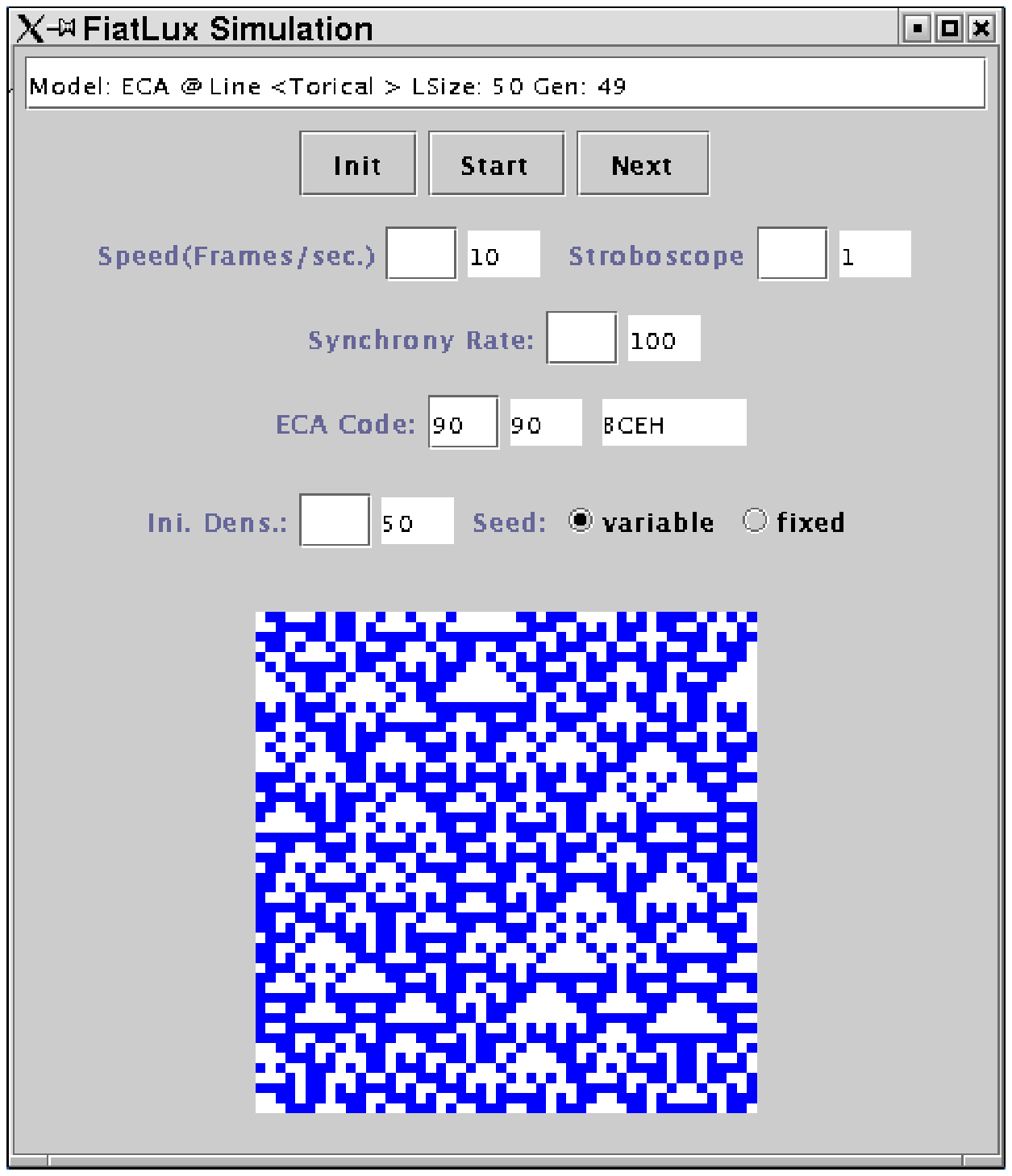}{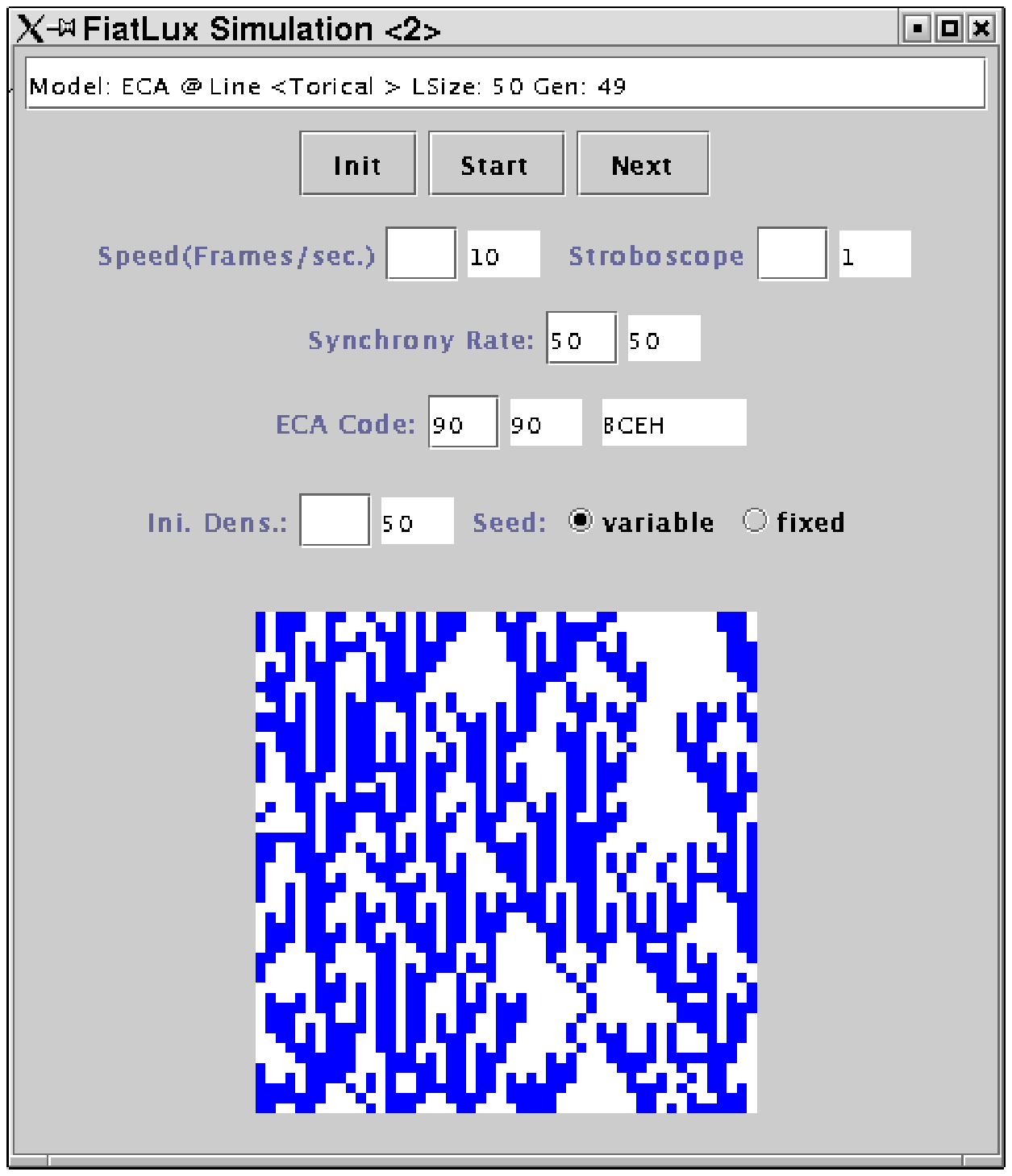}
\DoubleCaptionLabel{\LegSurfHorizontala}{\LegSurfHorizontalb}{ECA90lbl}
\EF

Rules such as ECA \ECA{90} and \ECA{150} have been among the most extensively studied rules of the ECA space. 
They are said to be 'additive' as they obey a superposition principle (see \ref{Sec:superposition}).
For these two rules, we found a horizontal sampling surface with $ \muf \eq 0.5 $ for all $ ( \dini, \sr )$.
This means that the qualitative behavior of the model is invariant when both changing the initial density and the synchrony rate.
Indeed, experimental evidence in the synchronous case shows that for any random initial density, 
the dynamical systems rapidly evolves towards an ``equilibrium state'' for which the density oscillates around $ \dens = 0.5 $ \cite{Wol83}.
In both synchronous and asynchronous case, 
this ``equilibrium state'' is not a fixed point but is rather a random phase 
in which the fluctuations of each cell appear to be random (see \figref{ECA90lbl}).
This implies that the model's robustness is explained by H1, 
more precisely, we expect the distribution of the density after the ``transient time''
to be a Gaussian with a mean centered around $ \dens = 0. 5$ and a variance that is proportional to
 $ 1/\sqrt{n}$~, where $ n $ is the lattice size. 
If this assumption is correct, then we have $ (\ra,\rb) \fleche (0,0) $ as $ \Tt \fleche \infty $ and $  \Ts \fleche \infty $,
which is what we observed experimentally when increasing $ \Tt $ and $ \Ts$.

\subsubsection{$ \dini $-dependent, $ \sr $-invariant surfaces}

\BF
\SSurf{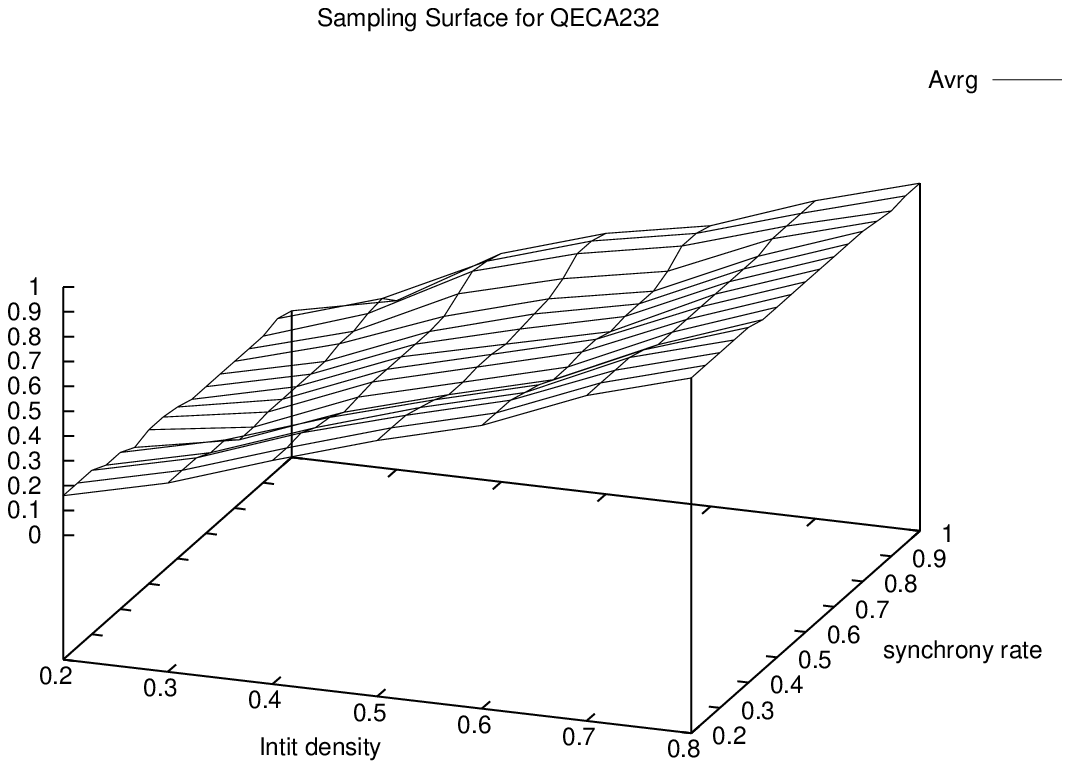}
\DoubleOrbit{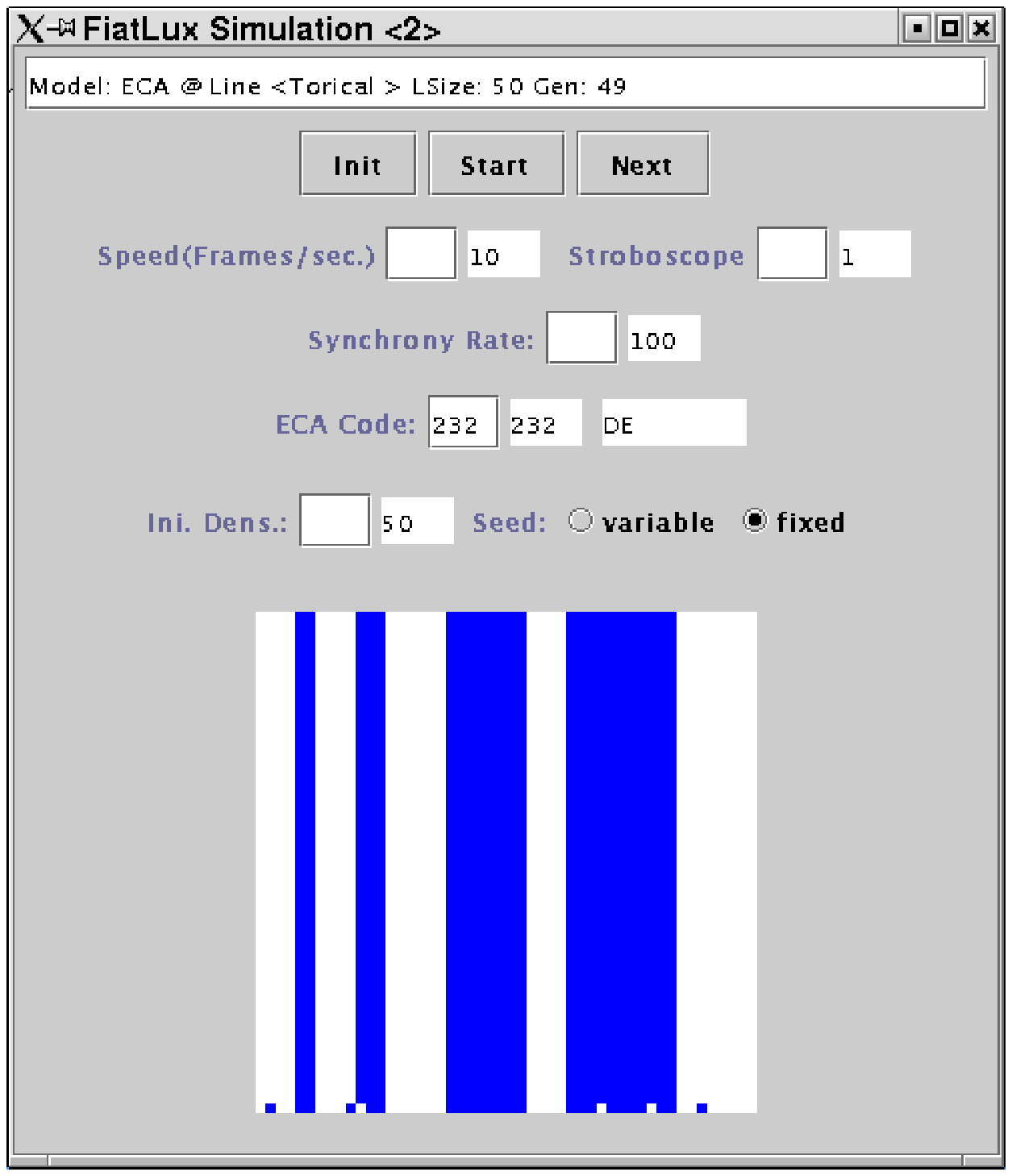}{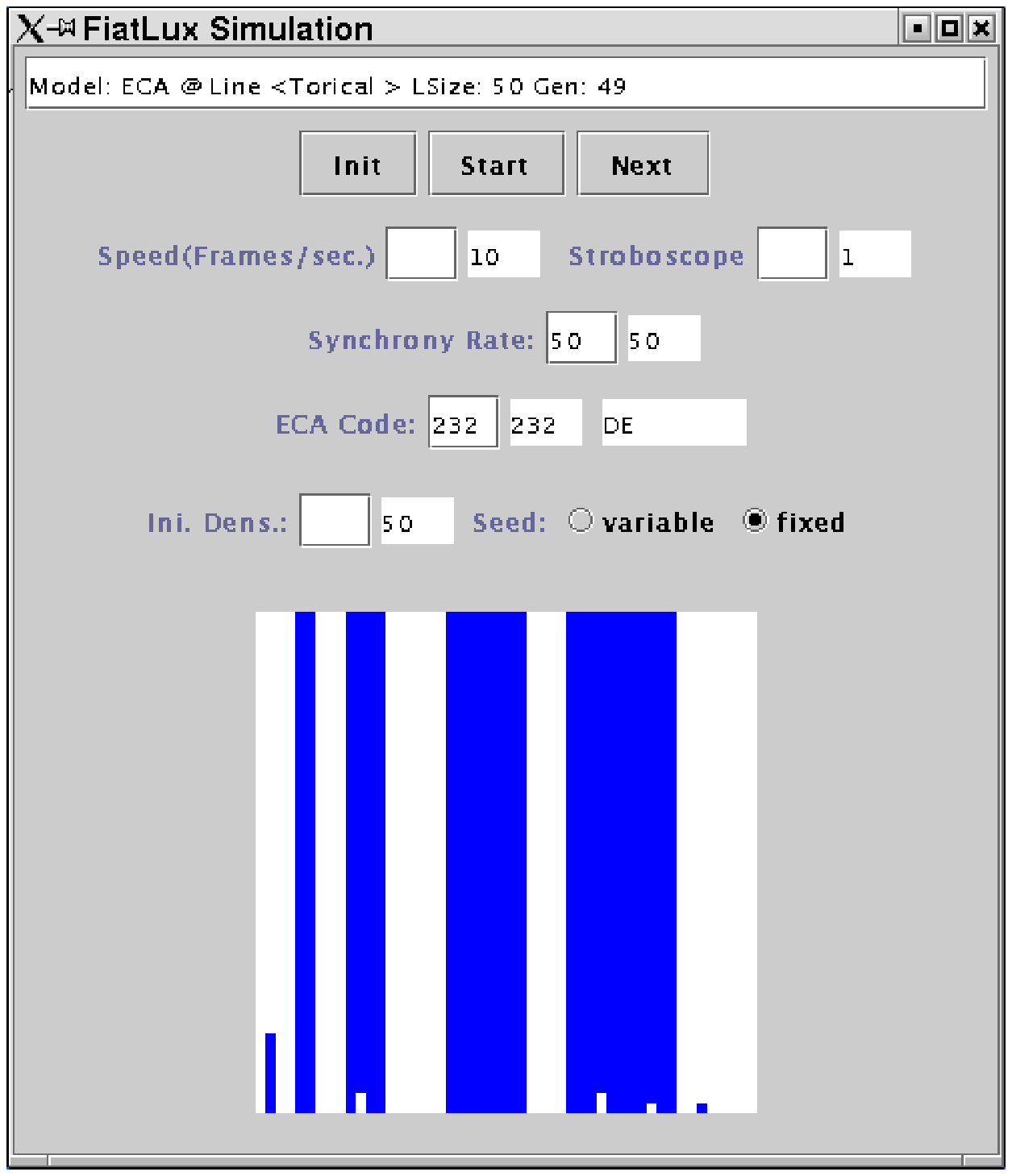}
\DoubleCaptionLabel{\LegSurfMaja}{\LegSurfMajb}{ECA232lbl}
\EF

ECA \ECA{232} is an ECA version of the ``Majority Vote Rule'' : the next state of a cell is the state that it is most present in its neighborhood. 
We found that this model is a good example of a Zone A ECA 
with a sampling surface that shows dependence on the initial density $ \dini $ 
and invariance with translation in the $ \sr $ axis~: see \figref{ECA232lbl}.
The dependence on $ \dini $  is explained by the existence of walls (\Word{00} and \Word{11}) for this rule.
These walls appear in the initial configuration or they are created when the dynamical system evolves and we observed a 
quick convergence of the orbits to a fixed point as seen in \figref{ECA232lbl}.
This convergence implies that the model's robustness is explained by H2 as the asymptotic part of the orbits is always a fixed point.

ECA \ECA{4}, \ECA{12}, \ECA{44}, \ECA{76} are some others zone A models which showed quick convergence to a fixed point. 
We can note that for all these models, the local transition rule admits walls\footnote{
\Word{0} and \Word{010} are walls of rule \ECA{4}, \Word{0} and \Word{01} are walls of rule \ECA{12}, 
\Word{00} and  \Word{0001}are walls of rule \ECA{44}, \Word{0},\Word{01}, \Word{10} are walls of rule \ECA{76}.}.
The question of knowing how the shape of sampling surface is related to the existence of walls
is a potential theoretical problem that arises from these observations and that should be addressed in the future.

\subsubsection{Perfectly $\sr$-invariant sampling surfaces}

Interestingly enough, the analysis of experimental data shows that some ECA are situated exactly on the point  $ (\ra, \rb) = (0, 0) $.  
Their sampling surface is thus perfectly invariant with translation in the $ \sr $ axis. 
This means that given a specific initial condition, 
the choice of the synchrony rate did not influence the value taken by the observation function $ \muf $.
The visual examination of the orbits of these particular ECA shows for a given initial condition, 
all orbits (for different $\sr$) converge to the {\em same} fixed point~:
\[ \forall x_i \in E, \exists x_f \in \E, \forall \sr \in ]0,1], \exists t, \gamma_{\sr}(x_i,t) = x_f \FP \]

We define the class of ``perfectly robust'' (PR) CA as the class of models for which 
the ``asymptotic behavior'' of a CA is independent of the updating method $ \dyn $, 
with $ \dyn $ verifying the fair sampling condition (see \ref{Sec:FairSampling}).
Some PR rules can be exhibited in a straightforward way.
For ECA \ECA{0} (null rule), as every cell update turns the cells into state 0, under the fair sampling condition, 
we are sure to reach the fixed point $ \zero $.
For ECA \ECA{204} (identity), any initial condition is a fixed point and the update does not play any role.
If we look at ECA \ECA{128} (see \figref{SPTfig}), all cells turn to state $ \stzero $ unless they are in state $ 1 $ and surrounded by two $ \stzero $. 
It is easy to see that the two only fixed points are $ \zero $ and $ \one $ and 
that any configuration different from $ \one $ evolves to the fixed point $ \zero $.

Experimentally, we find that~: PR= \set{ 0, 8, 32, 40, 128, 136, 140, 160, 168, 200, 204 (Identity) }. 

To find a sufficient and necessary condition to be in PR is another problem that arises from the analysis of the experimental results.

\subsection{Zone B (big $ \ra $, small $ \rb $)}

This zone contains the ECA for which a small introduction of asynchronism produces a brutal change of behavior (big $ \ra $), 
while this behavior no longer changes when asynchronism is increased (small $ \rb $).

\subsubsection{Surfaces with a discontinuity at $ \sr = 1 $ and flatness for the rest of the surface}

\BF
\SSurf{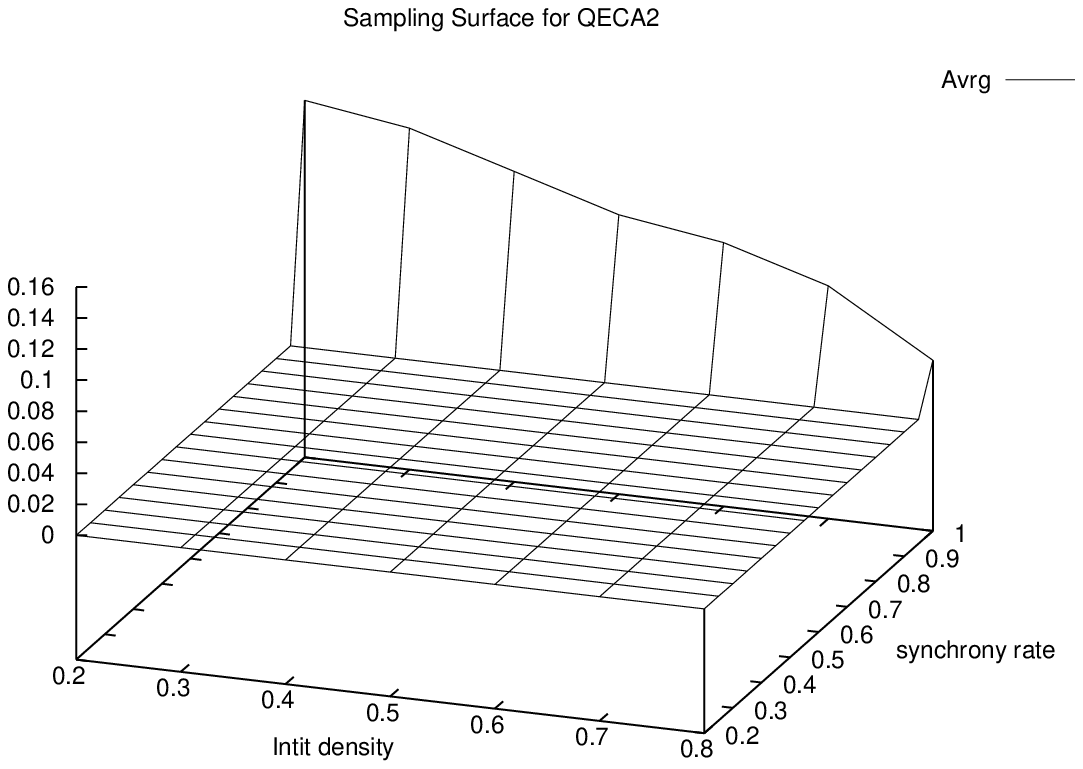}
\DoubleOrbit{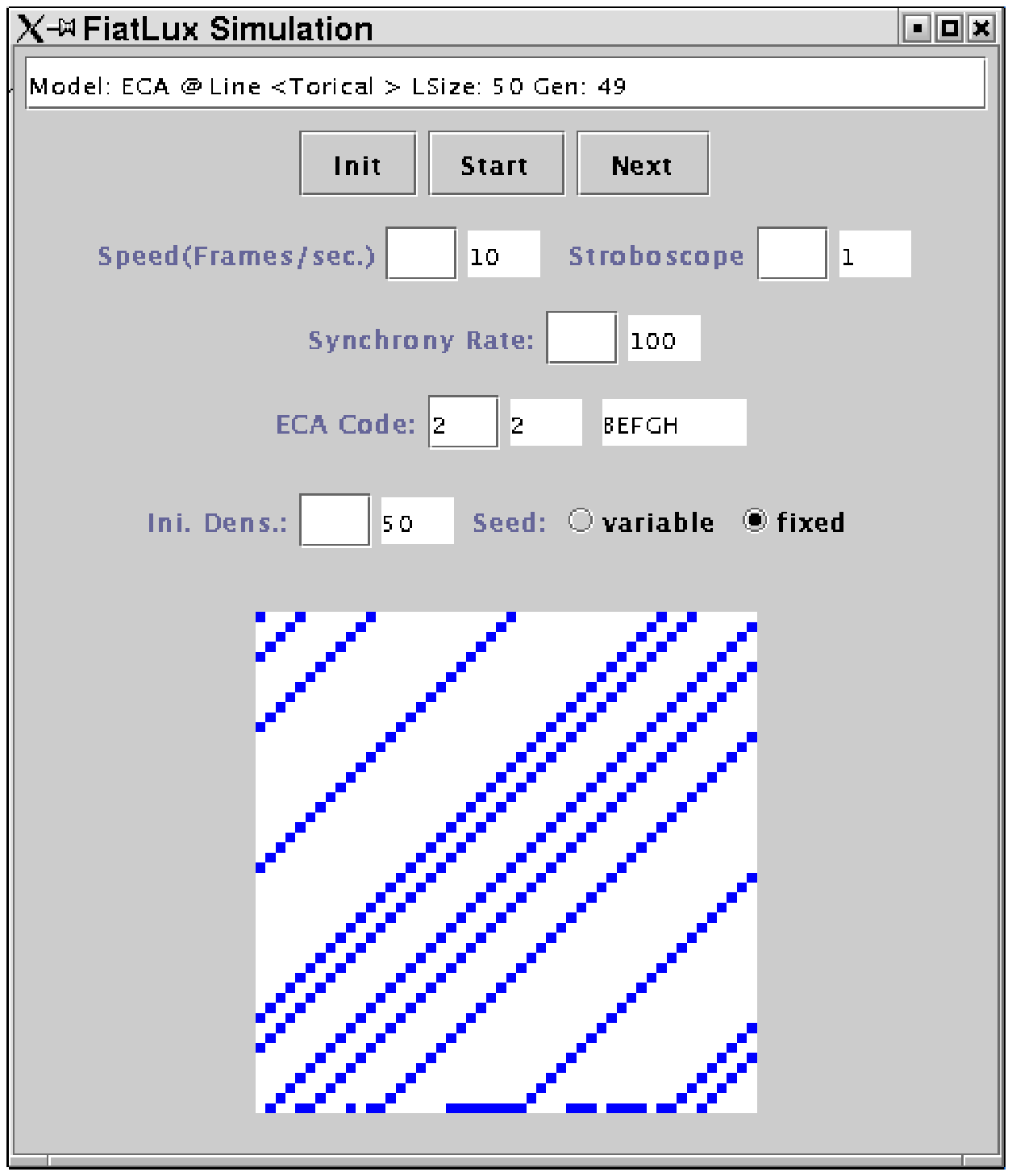}{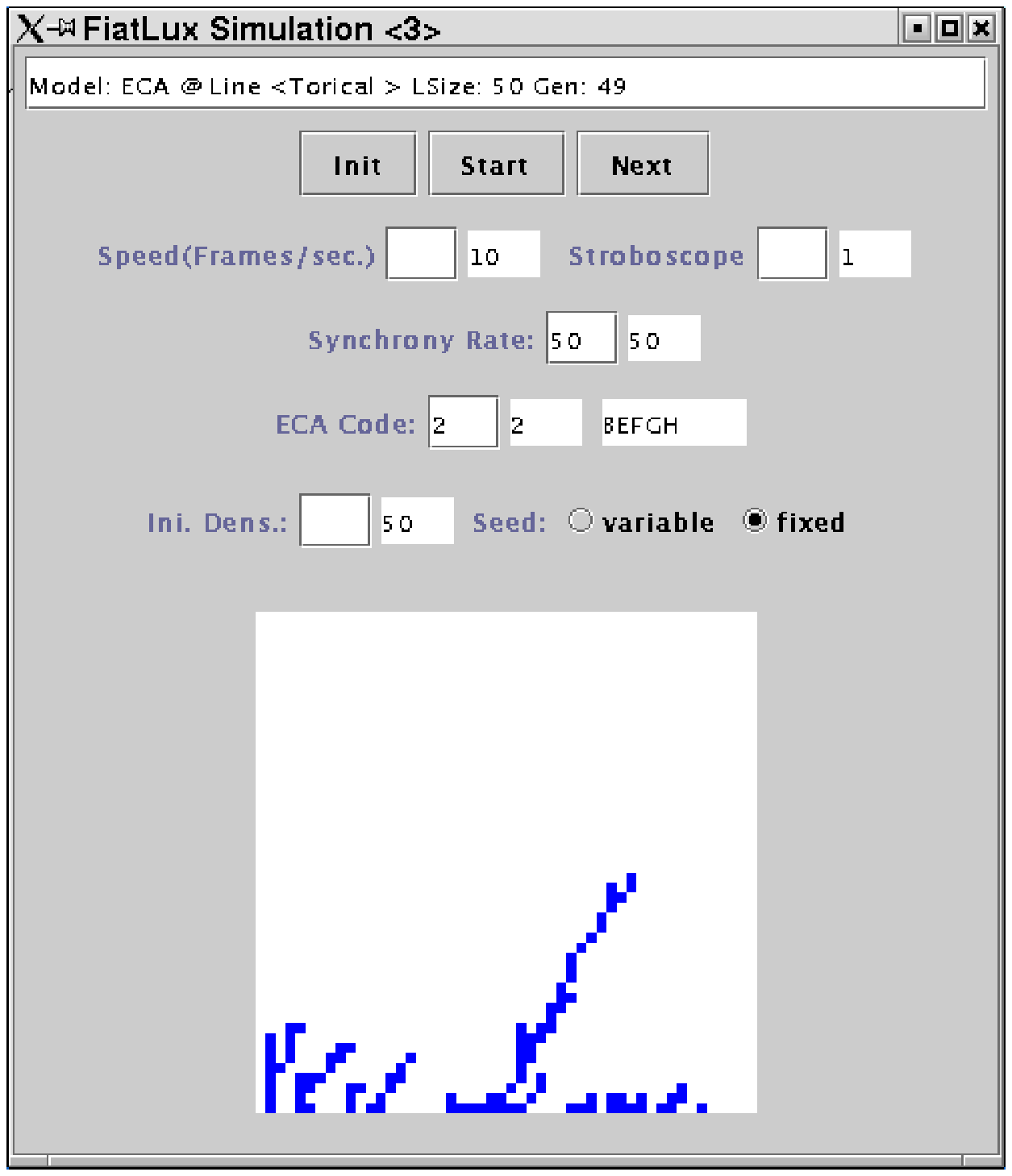}
\DoubleCaptionLabel{\LegSurfBIFZa}{\LegSurfBIFZb}{ECA2lbl}
\EF

In this zone, we can distinguish some ECA for which we have exactly $ \rb = 0 $.
Visual examination of the sampling surface shows that 
these CA exhibit a discontinuity of the surface, indicating a ``phase transition'' phenomenon, for the points $ \sr = 1 $.
When looking at the orbits of these ECA (see \figref{ECA2lbl}), 
we notice that for $ \sr = 1 $, the orbits evolve into a shift-like behavior, 
where each configuration gets translated by one cell at each time step.
For $ \sr < 1 $, the orbits evolve in similar way, except that some ``branches'' ($ \stone $-domains) progressively die out. 
This means that the orbit finally reaches a spatially homogeneous fixed point consisting in all $ \stzero $ (the configuration $ \zero $  ).

We define GAP as the class of models for which there is a gap in the sampling surface between the values for $ \sr = 1 $ and $ \sr < 1 $ 
whereas the sampling surface is perfectly horizontal for $ \sr < 1 $. 
Experimentally, we find that GAP= \set{2, 10, 24, 34, 42, 56, 74, 130, 154, 162}.

We notice that all ECA in class GAP are ``fully asymmetric'' (i.e., there are four members in each equivalence class).
Moreover, all these rules except \ECA{154} are classified as ``subshifts'' by Cattaneo and al.\cite{Cat99}\footnote{
ECA \ECA{154} is symmetric to rule \ECA{180} which has been extensively studied in \cite{Cat98} 
where it was classified as a ``generalized subshift'' rule. In the classification proposed in \cite{Fates03}, 
the particular behavior of this rule was also noticed as \ECA{154} was classified in the ``hybrid'' (H) class.
}.
The asymmetry to the left/right exchange symmetry indicates that the rule has an isotropy which allows a directed propagation 
of some subwords to happen thus allowing the ``subshift'' phenomenon in the synchronous mode. 
On the other hand, the asymmetry to the $ \stzero $/$ \stone $ complementation shows that the rules may have a ``favorite'' state to which to tend to, 
thus explaining why the attractor $ \zero $ is reached with all the sampled initial conditions in the asynchronous regime.

\subsubsection{Surfaces showing a ``phase transition'' at $ \sr = 1 $ and quasi-flatness elsewhere }

\BF
\SSurf{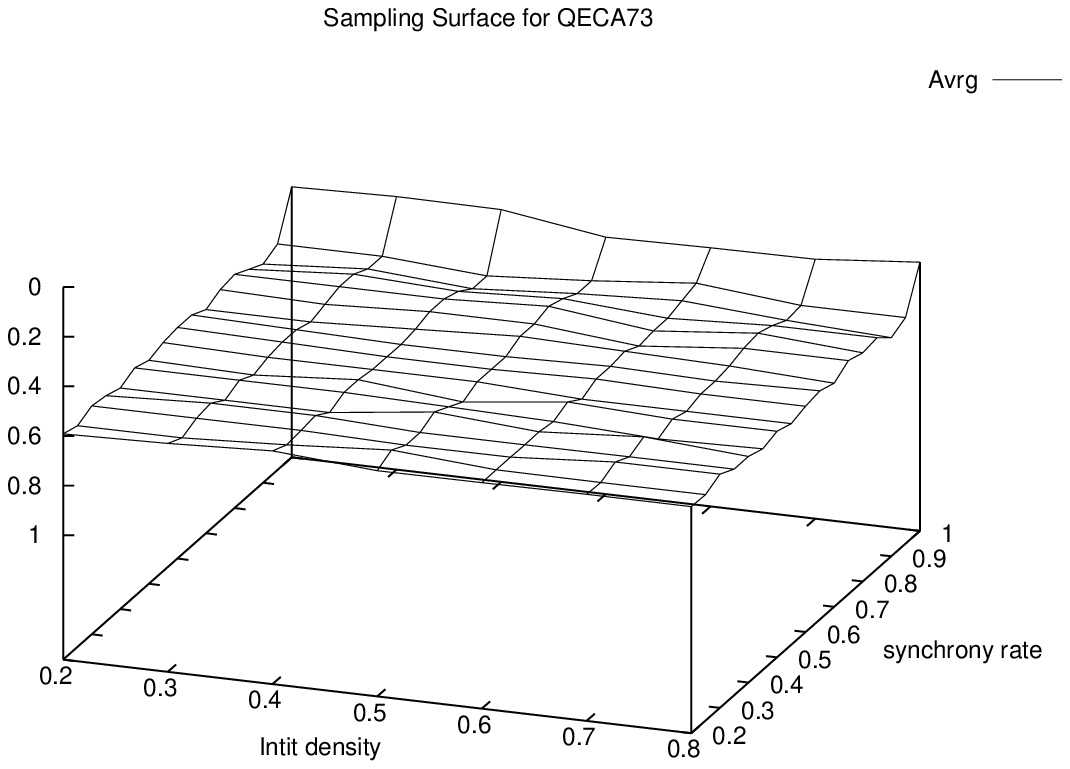}
\DoubleOrbit{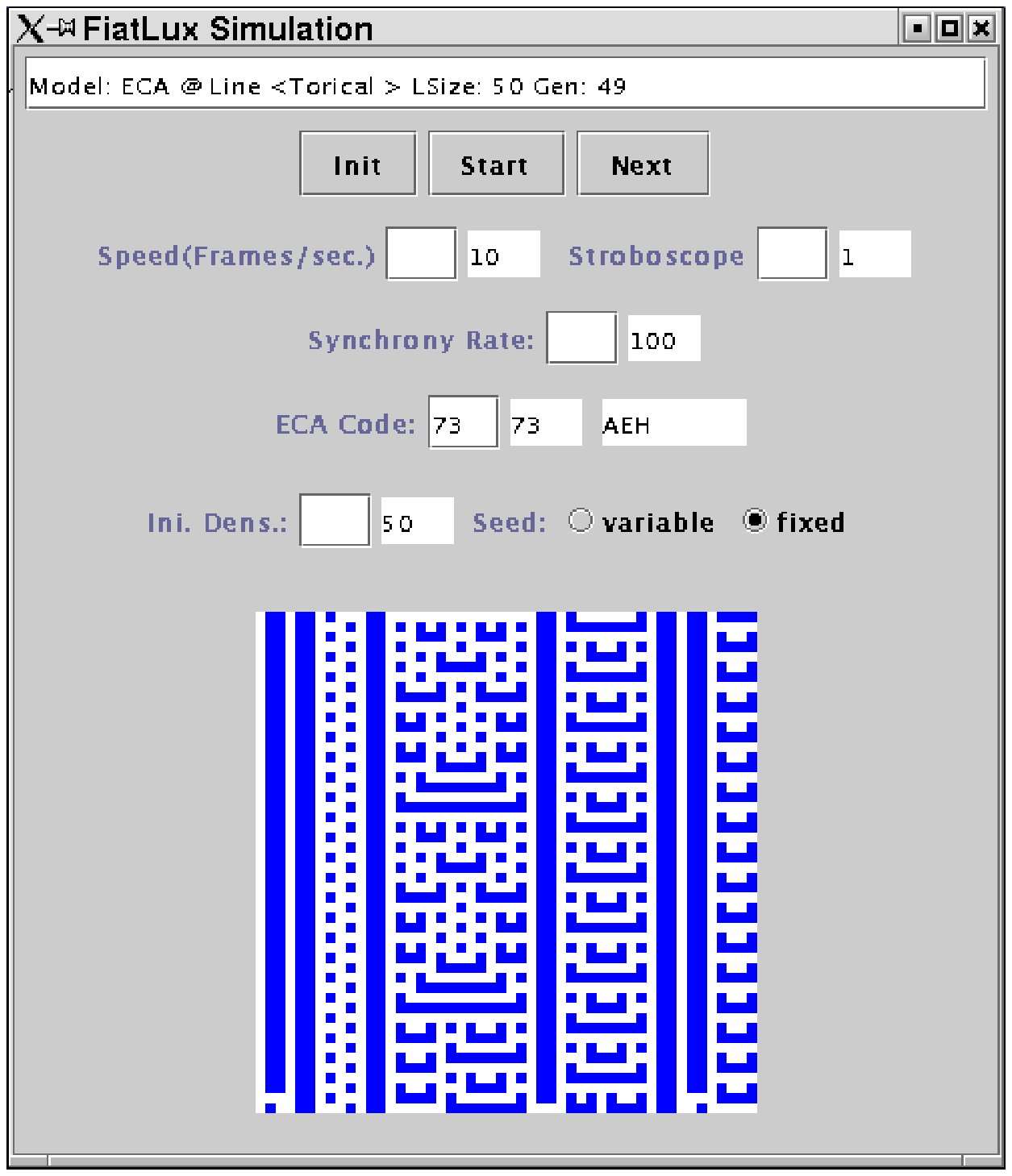}{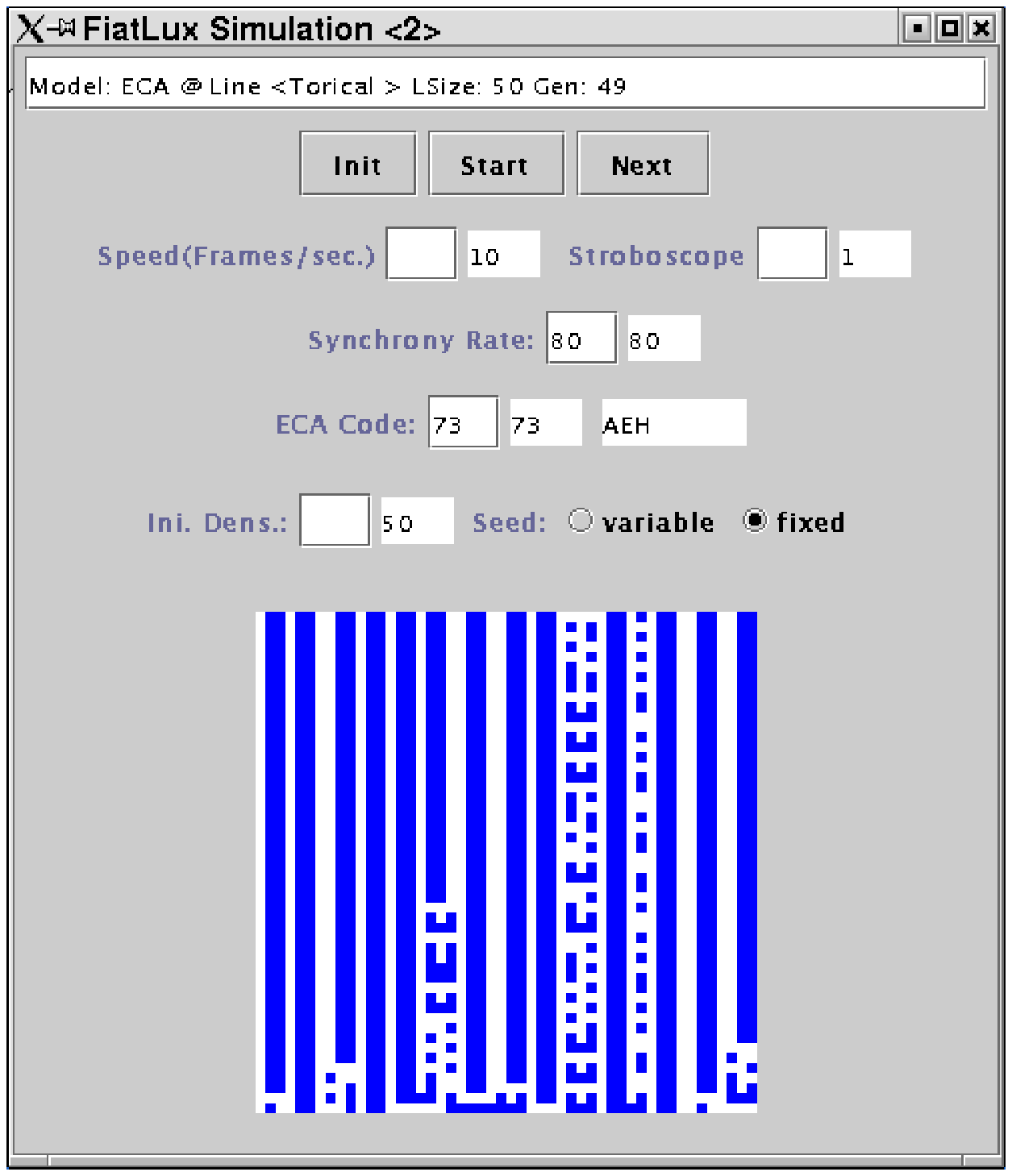}
\DoubleCaptionLabel{\LegSurfBifZNa}{\LegSurfBifZNb}{ECA73lbl}
\EF

For rules \ECA{73} and \ECA{142}, the examination of their sampling surface (\figref{ECA73lbl}) showed that 
an important change of the value of the observation function $ \muf $ occurs for $ \sr = 1 $. 
On the other hand, in the asynchronous part ($ \sr < 1 $), the surface appears flat though affected by a little irregularity.

The shape of the surfaces can be explained by the examination of the orbits of the models. 
As far as the dynamics is concerned, \ECA{73} is a border line CA~: 
visual examination of its orbits  (see \figref{ECA73lbl}) can not clearly help to decide 
whether it is in Wolfram's class II (periodic ECA) or in class III (``chaotic'' or non-regular ECA) \cite{Wol84}.
It is a ``Hybrid'' (class H) rule according to the classification exposed in \cite{Fates03}. 
Indeed, when evolved with perfect synchrony the model has a dynamics that is chaotic-like in some parts of the configuration 
delimited by walls $ \Word{0110} $. 
When a little asynchrony is introduced, there is a non-zero probability that a wall \Word{0110} appears in $\stzero$-domains 
where it was not already present.
This means that, as time progresses, more walls appear and the orbit eventually reaches a ``quasi-stable state'' 
in which the walls \Word{0110} are separated by three kind of subwords~:
\BL
\item $ \Word{0} $ ~: these subwords are stable
\item $ \Word{00} $ ~: these subwords are stable
\item $ \Word{000}$ and $  \Word{010} $~: theses two subwords alternate one after another 
	when the update rule is applied in the middle of the word.
\EL

This quick analysis allow us to understand the shape of the sampling surface : 
the first gap showed by the observation function is due to the appearance of walls when little asynchronism is added, 
the fluctuations in the surface are due to the random updatings of the $ \Word{000}$ and $  \Word{010} $ regions.
ECA \ECA{73} and \ECA{142} are the only two elements found in Zone B and that do not belong to class GAP.

\subsection{Zone C (small $ \ra $, big $ \rb $) }

\BF
\SSurf{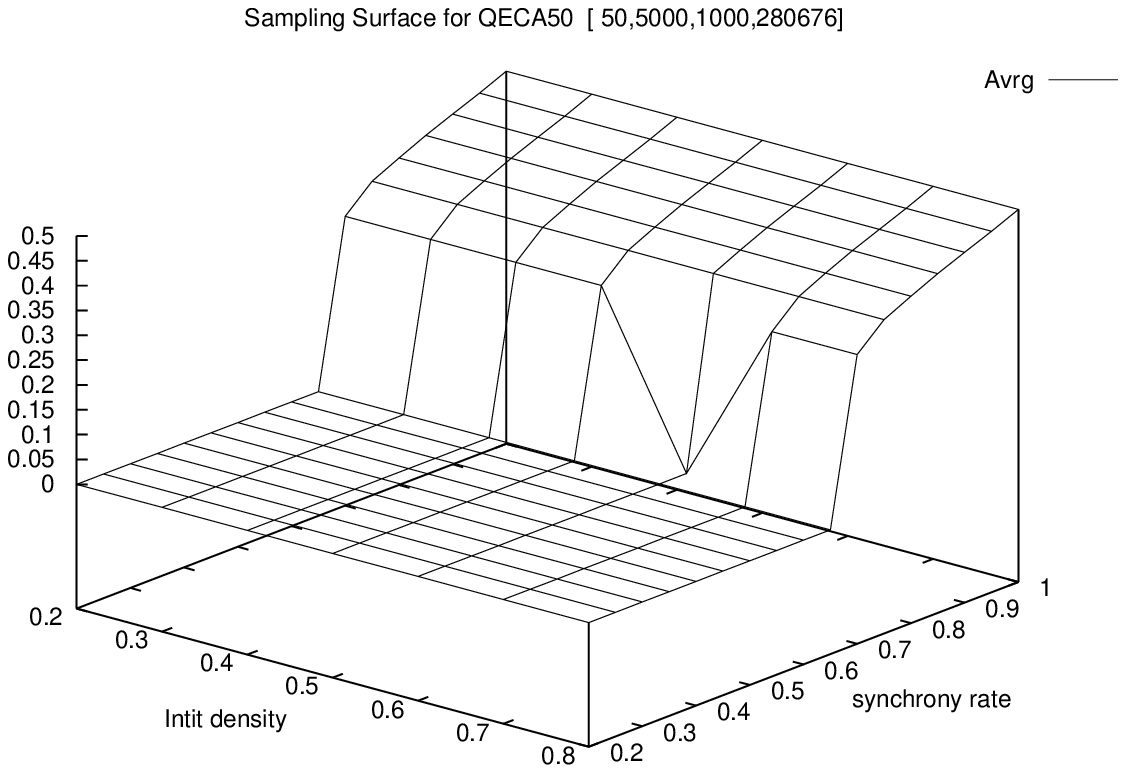}
\TripleOrbit{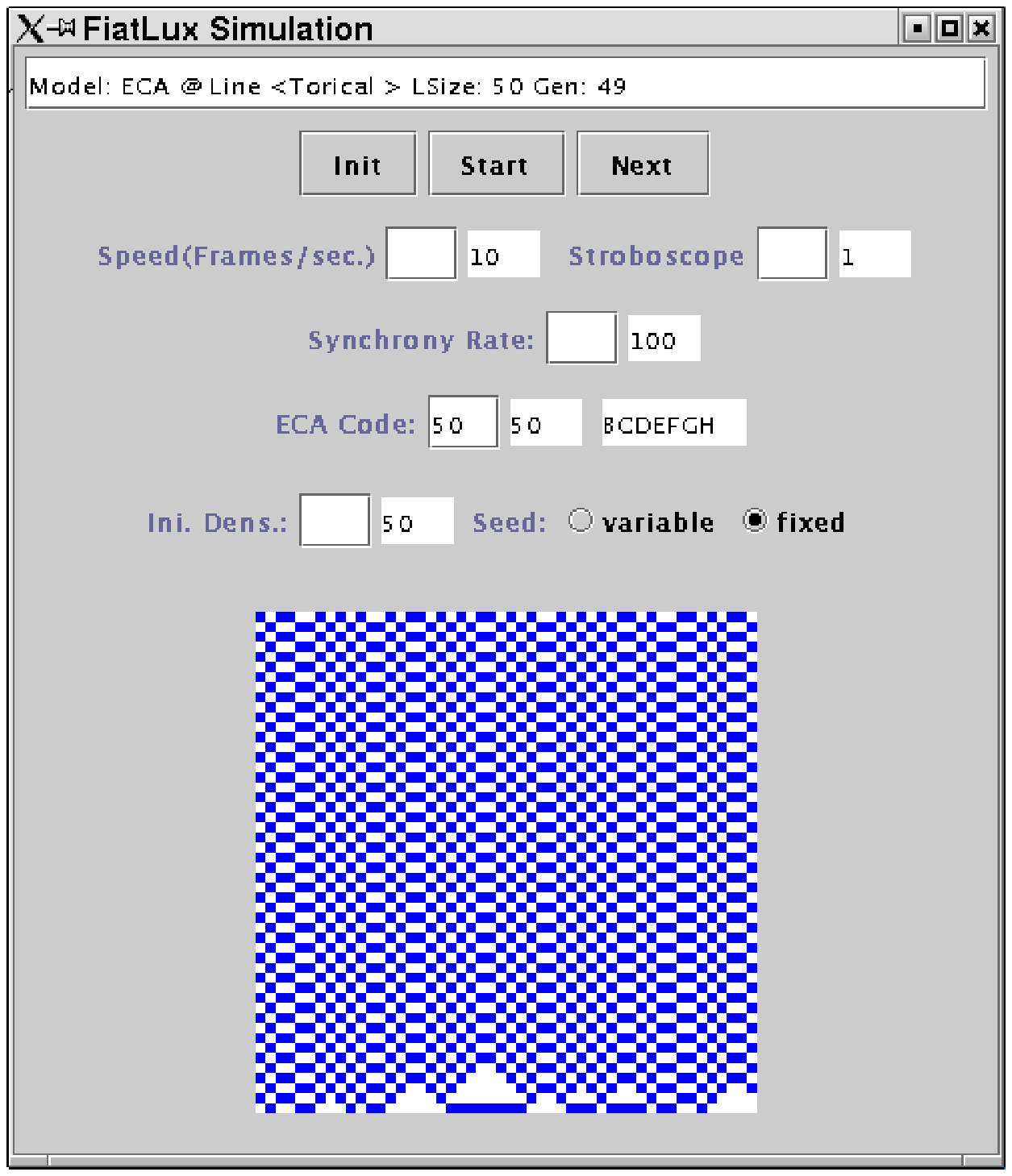}{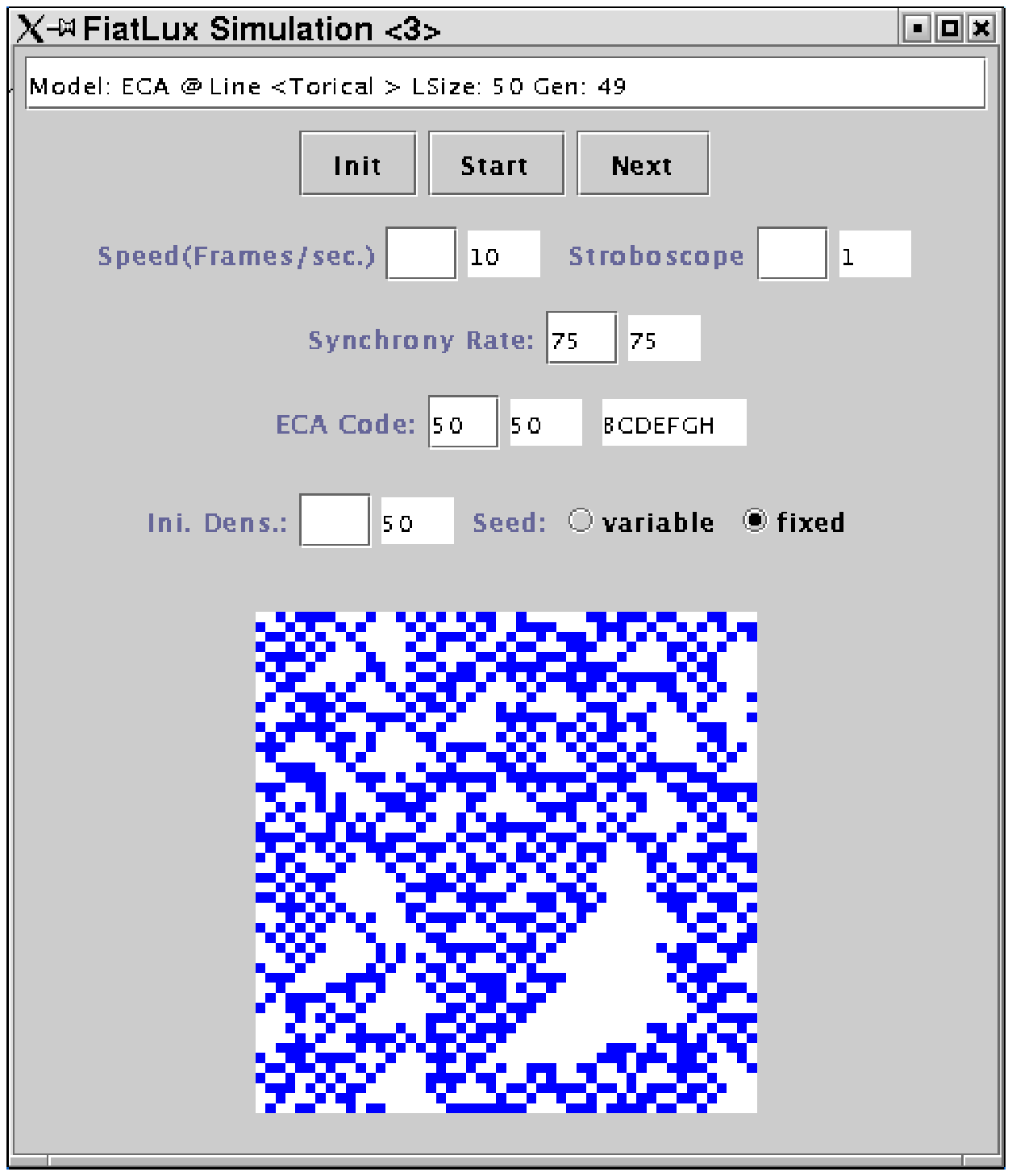}{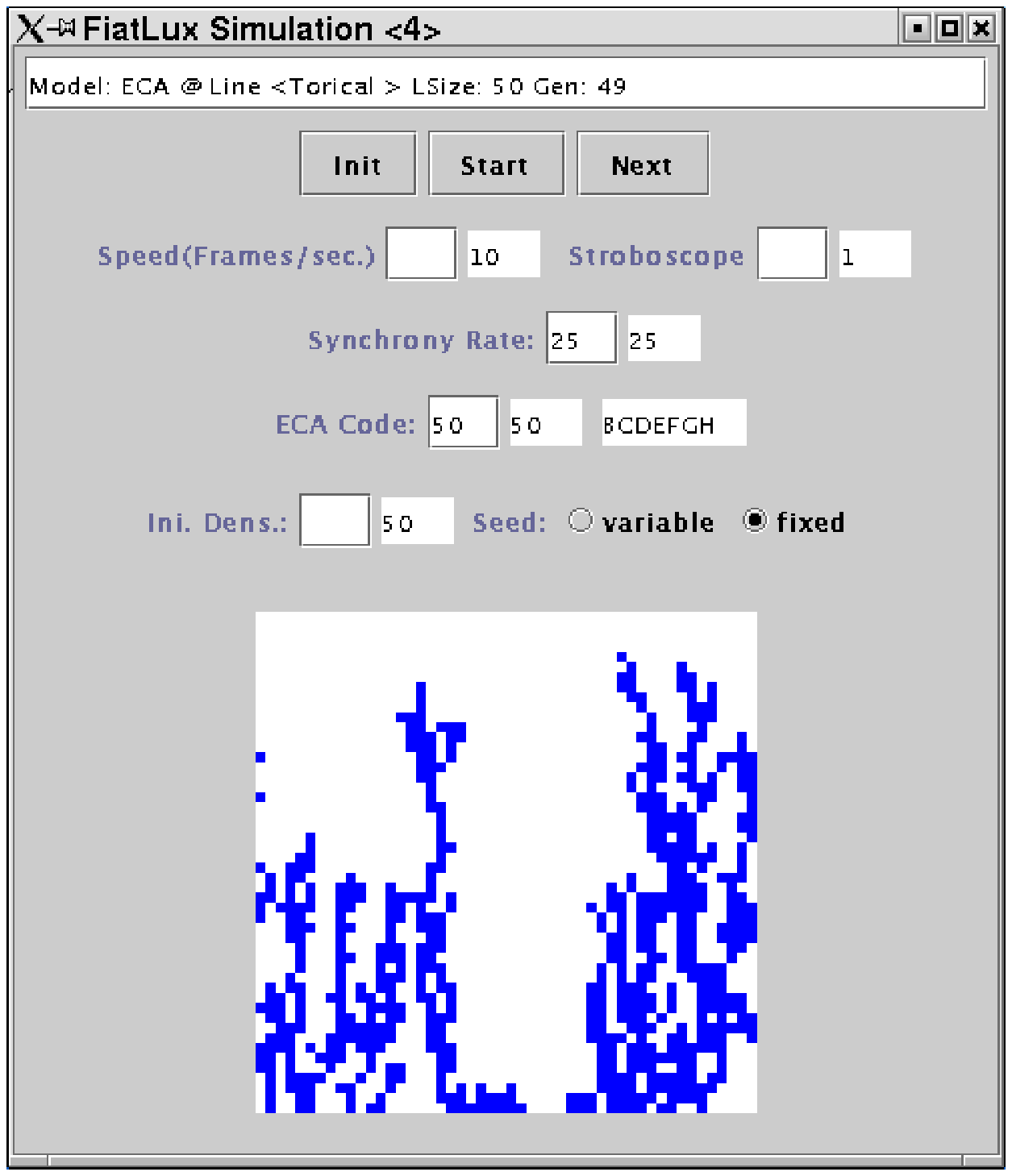}
\DoubleCaptionLabel{\LegSurfBIFOa}{\LegSurfBIFOb }{ECA50lbl}
\EF

In this zone, we find the ECA for which an important change of behavior occurs for values of synchrony rate $ \sr < 1 $.

\subsubsection{Surfaces showing a ``phase transition'' at $ \sr_c < 1 $ }

In Zone C, we find some ECA with a sampling surface which clearly exhibits a discontinuity for a particular value of $ \sr_c $.
We have regrouped this type of models in the class SPT (Single Phase Transition).

The analysis of the orbits (see \figref{ECA50lbl}) of SPT members showed that 
for synchrony rates $ \sr > \sr_c  $, the evolution of the space-time diagram can be described in terms
of branching structures formed of  $ \stone $-domains that evolve on a background of $ \stzero $. 
On the other hand, for synchrony rates $ \sr < \sr_c $, the branching structure quickly dies out and 
the orbit reaches the fixed point $ \zero $.
This kind of phenomenon has already been noticed in the study of coupled map lattices and an analogy was made with fluid mechanics : 
the turbulent phase is represented by the branching structure and the laminar phase is represented by the background of $ \stzero $ (absorbing state). 
The laminar phase is stable and can only be destabilized by the diffusion of the turbulent phase.
For continuous-state systems, it has been conjectured that the phenomenon of branching structures could be described in terms of directed percolation \cite{Pomeau86}. We are at the moment unable to provide a suitable description for the discrete models, even though
the work of Chat\'e and Manneville showed that some insight could be gained 
by understanding CA behavior in terms of discretized coupled map lattices \cite{Chate88}, \cite{Chate90}.

Experimentally, we find that~: SPT= \set{ 6, 18, 26, 50, 58, 106, 146, 178 }.
 
Note that \ECA{22} and \ECA{30} have a similar ``phase-transition'' 
behavior : in this case, the branching pattern is constituted of defaults of regularity of the regular background $ \Word{01} $.
This implies that the density of the orbits fluctuates near $ \dens = 0.5 $ and that the sampling surfaces are flat 
and do not allow to detect the qualitative change. 
ECA \ECA{178} has a parameter $ \rb $ that is much bigger than other SPT members (see \figref{EvolRep}).
This can be explained by the fact that it is the only member which has two attractors in the stable ``phase'' ($\zero$ and $\one$), 
thus producing higher potential changes between the stable phase and the unstable phase.

\subsubsection{$ \dini $-invariant,$ \sr $-dependent surfaces}

\BF
\SSurf{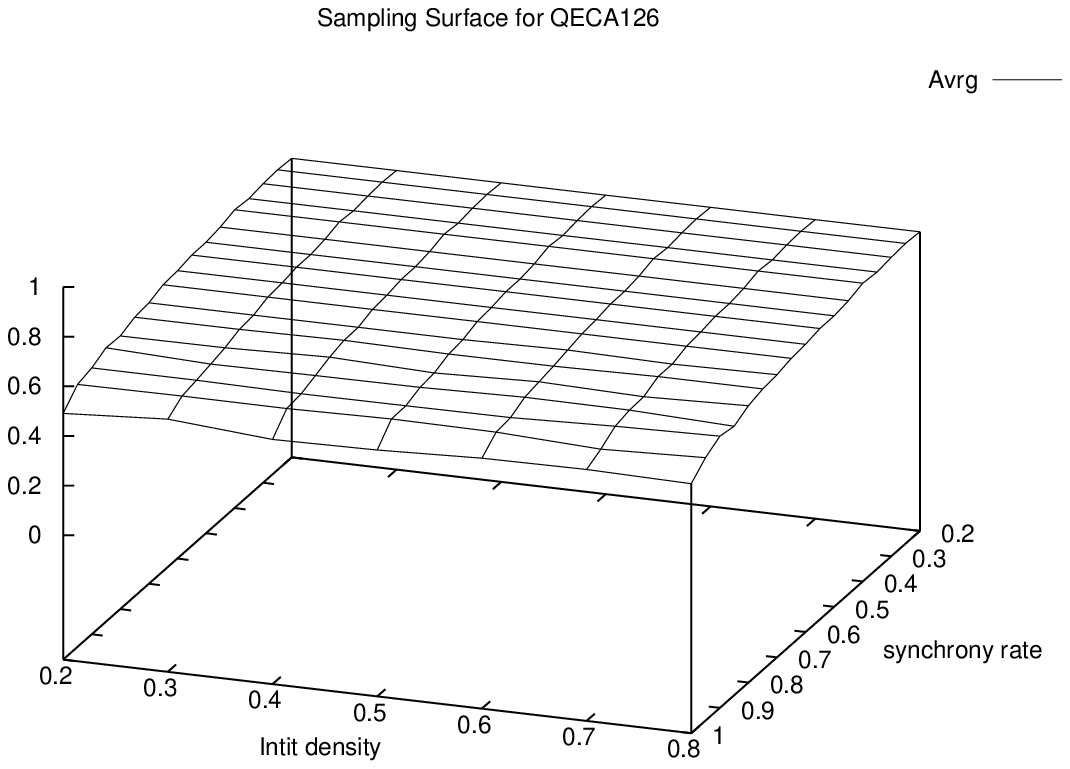}
\TripleOrbit{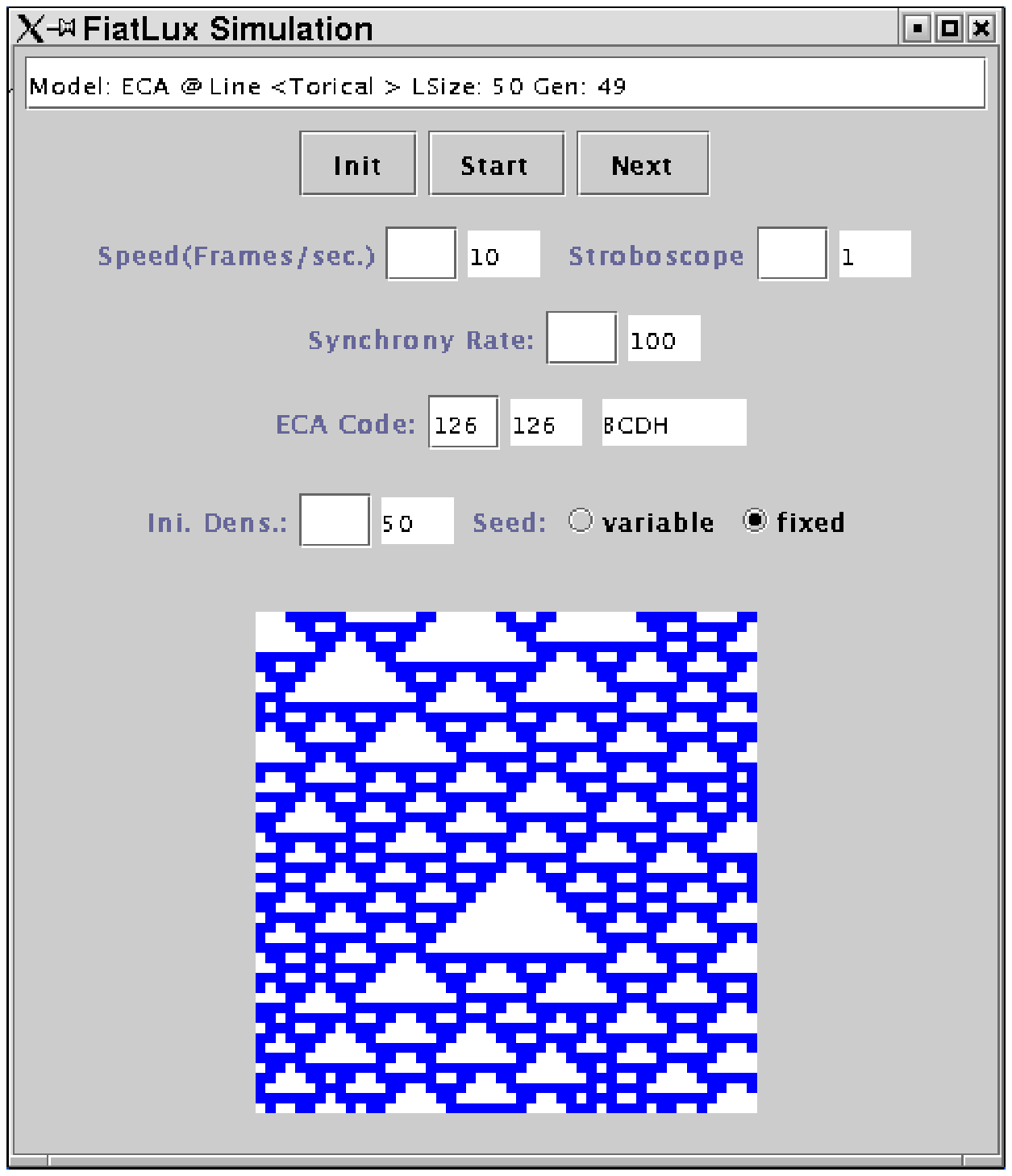}{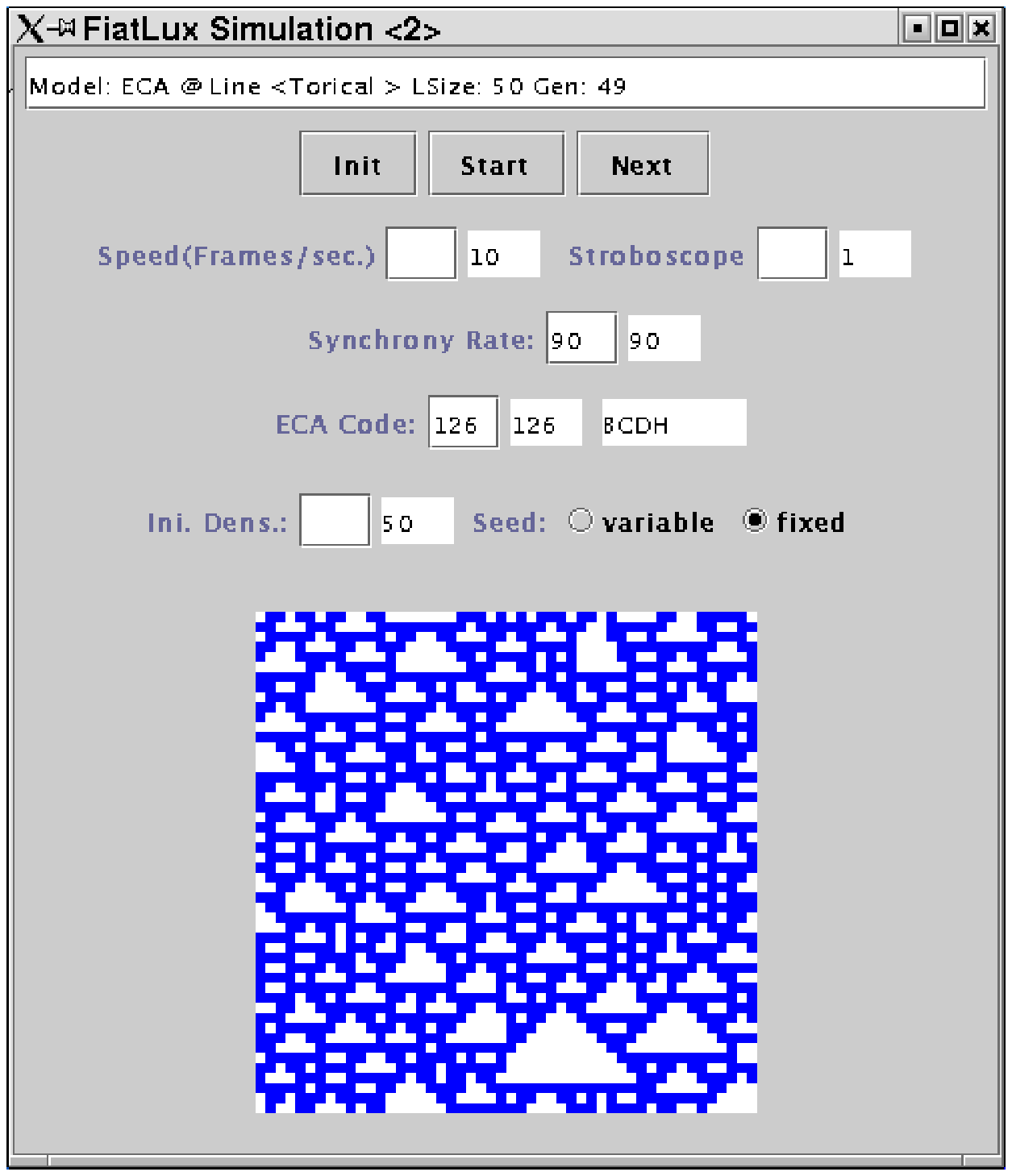}{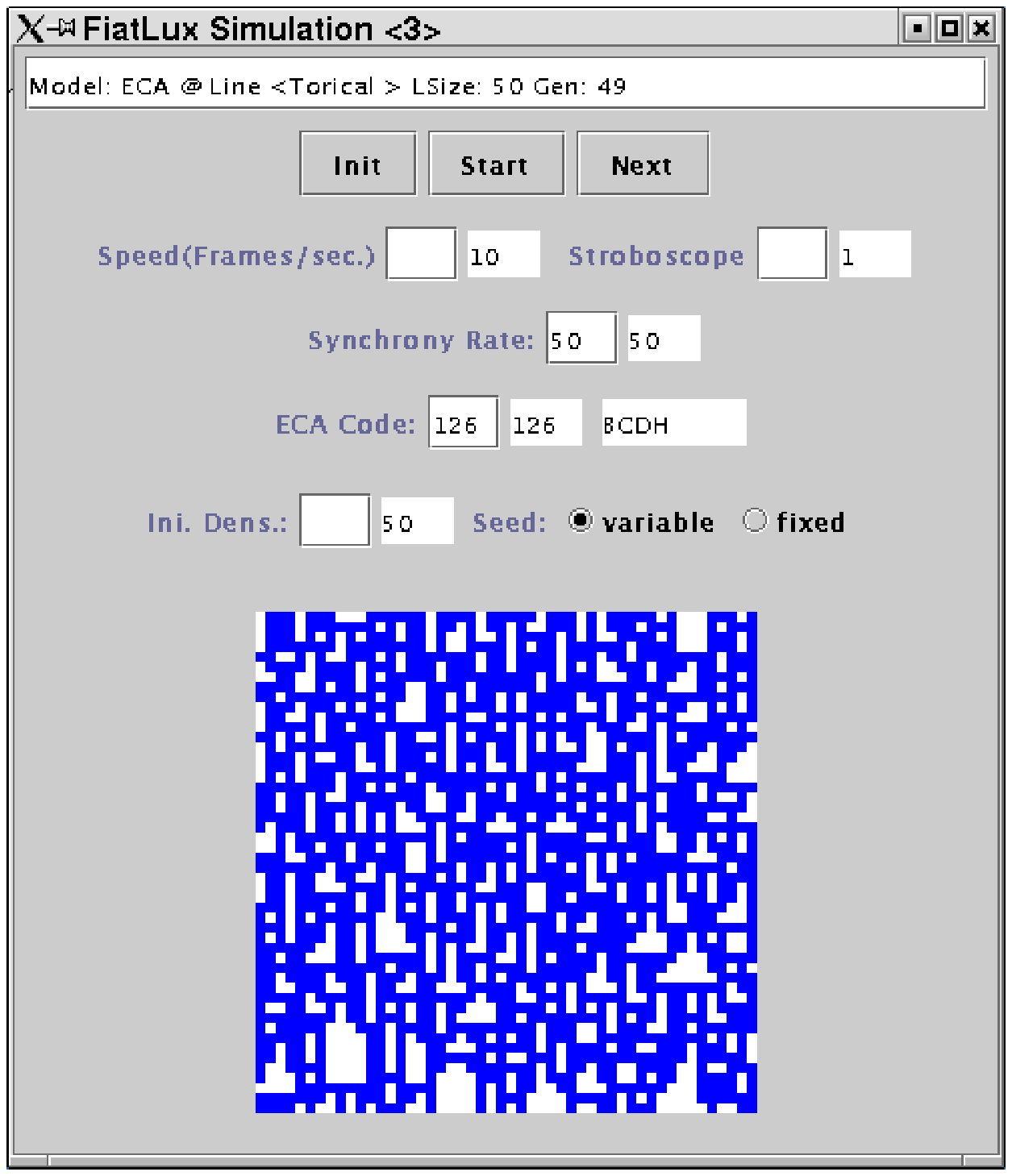}
\DoubleCaptionLabel{\LegSurfContia}{\LegSurfContib}{ECA126lbl}
\EF

We found that only \ECA{126} was in Zone C but not in SPT.
\ECA{126} is a class III CA (\cite{Wol84}) for which the evolution of the synchrony rate does affect the evolution of the density ``smoothly''
(see \figref{ECA126lbl}).

\subsection{Zone D (big $ \ra $, big $ \rb $ )}

\subsubsection{Unstable surfaces}

\label{Sec:ECA170}

\BF
\SSurf{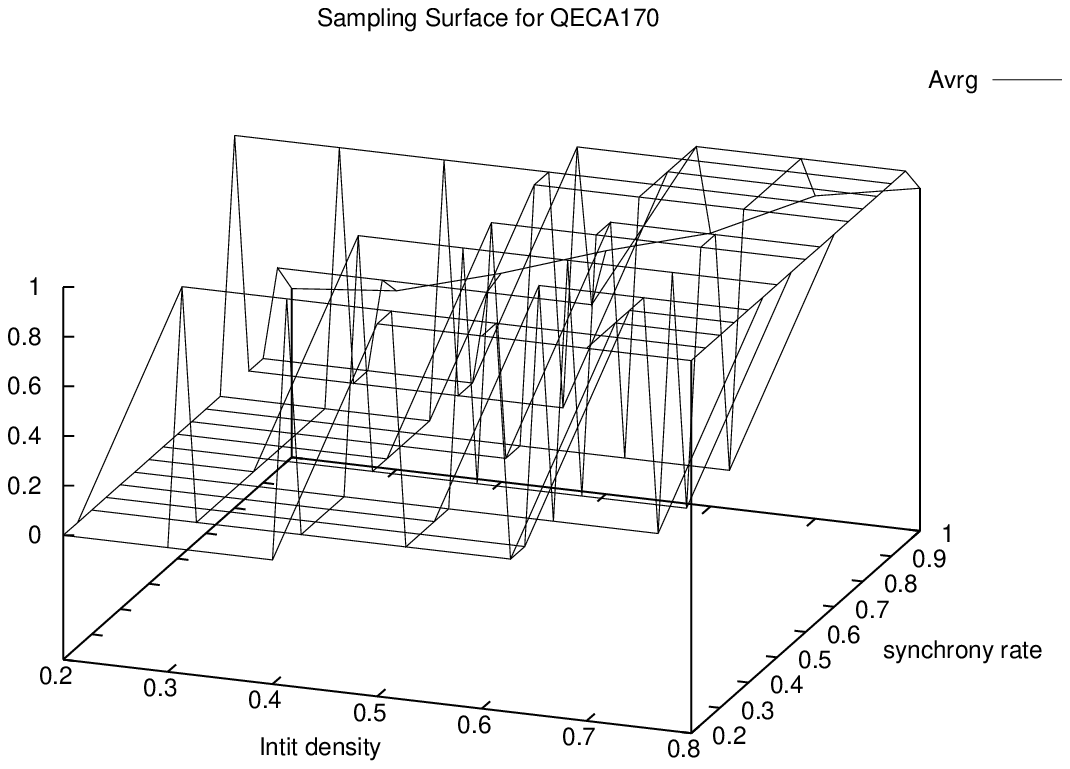}
\DoubleOrbit{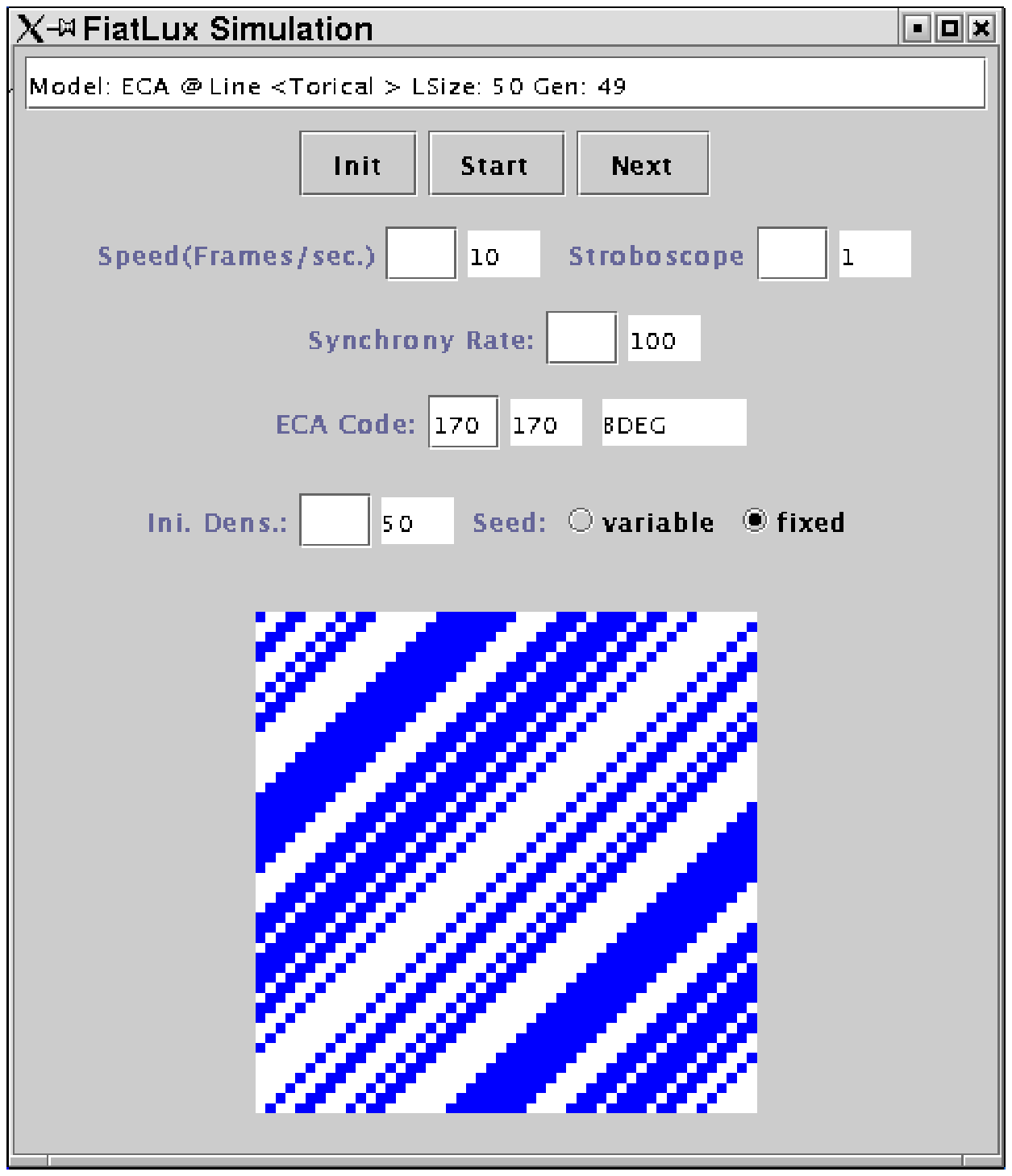}{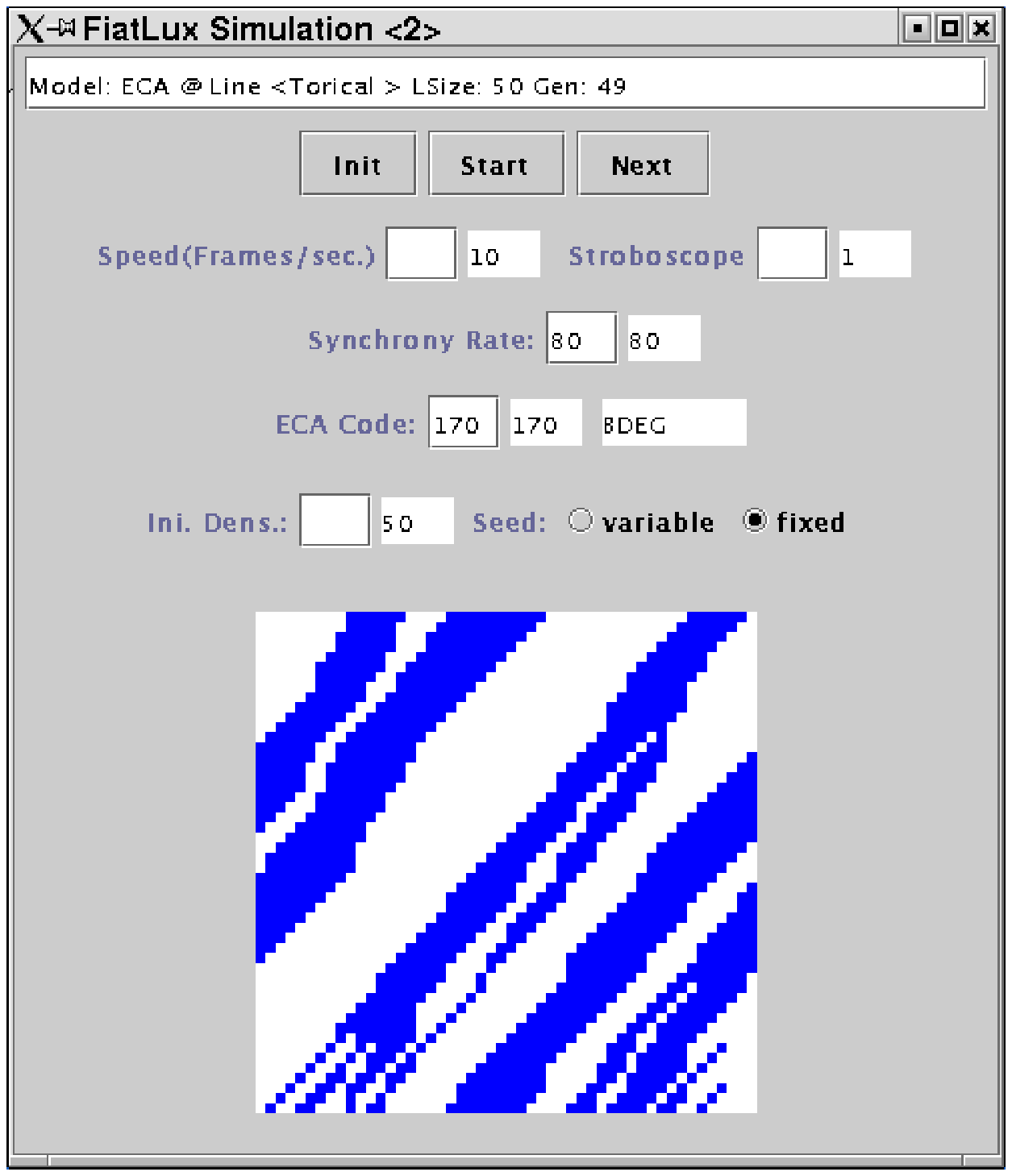}
\DoubleCaptionLabel{\LegSurfShifta}{\LegSurfShiftb}{ECA170lbl}
\EF

In this zone, we find the ECA for which the measure of $ \muf $ is highly unstable. 
When $ \ra $ and $\rb $ are high, 
this can indicate a bad statistical convergence of the parameters leading to the formation of a non-regular surface (see \figref{ECA170lbl}).
In these rules, when starting from any initial configuration different from  $ \zero $ or $ \one $,
we see that large zones of 0's or 1's appear and the borders of these zones drift in random way until they meet and annihilate. 
This is the case for ECA \ECA{138}, \ECA{170}(shift) and \ECA{184}. 

We notice that ECA \ECA{170} and \ECA{184} are two (non-trivial) number-conserving ECA in the synchronous case and this suggests that analytical
results could be obtained for such simple systems. 
ECA \ECA{138} is a rule which behavior is similar to \ECA{170} with one single difference on the output of the transition function~:
For $ (a,b,c) \neq (1,0,1)$ $ f( a, b, c) = a $ and $ f( 1, 0, 1) = 0 $,
this implies that the attractor $ \one $ is unreachable as a consecutive zone of $ \stzero $ can not disappear.

\subsubsection{A Sampling Surface with riddles~: ECA \ECA{46} }

\BF
\SSurf{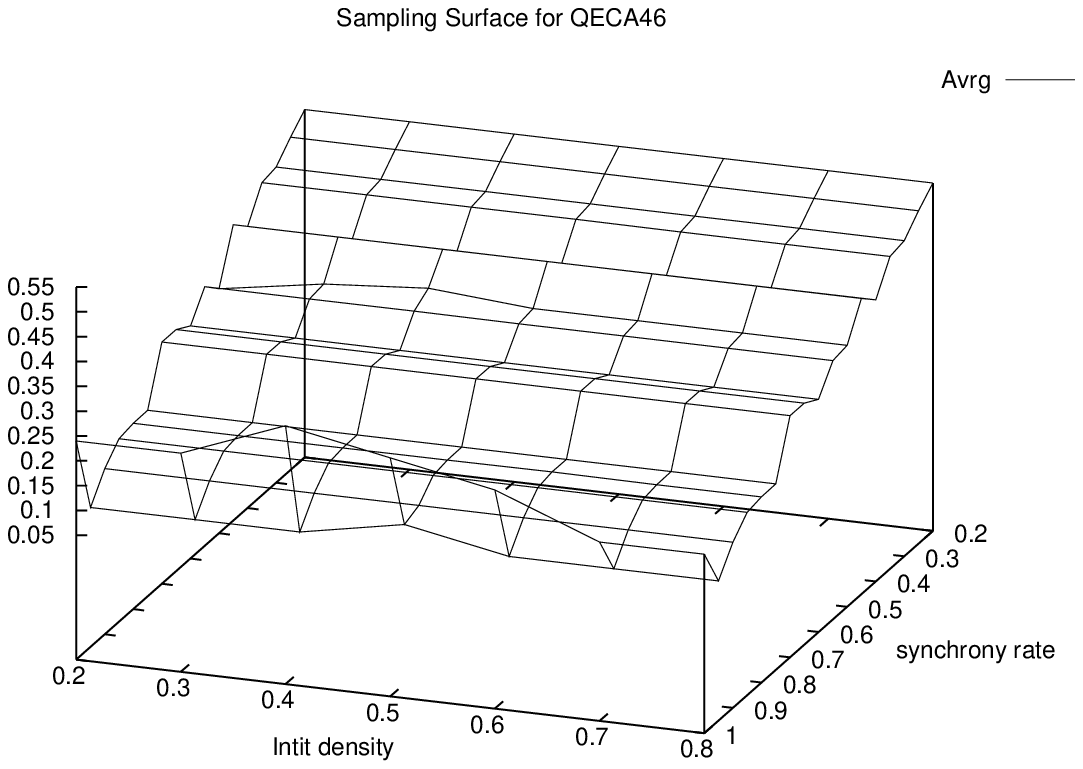}
\TripleOrbit{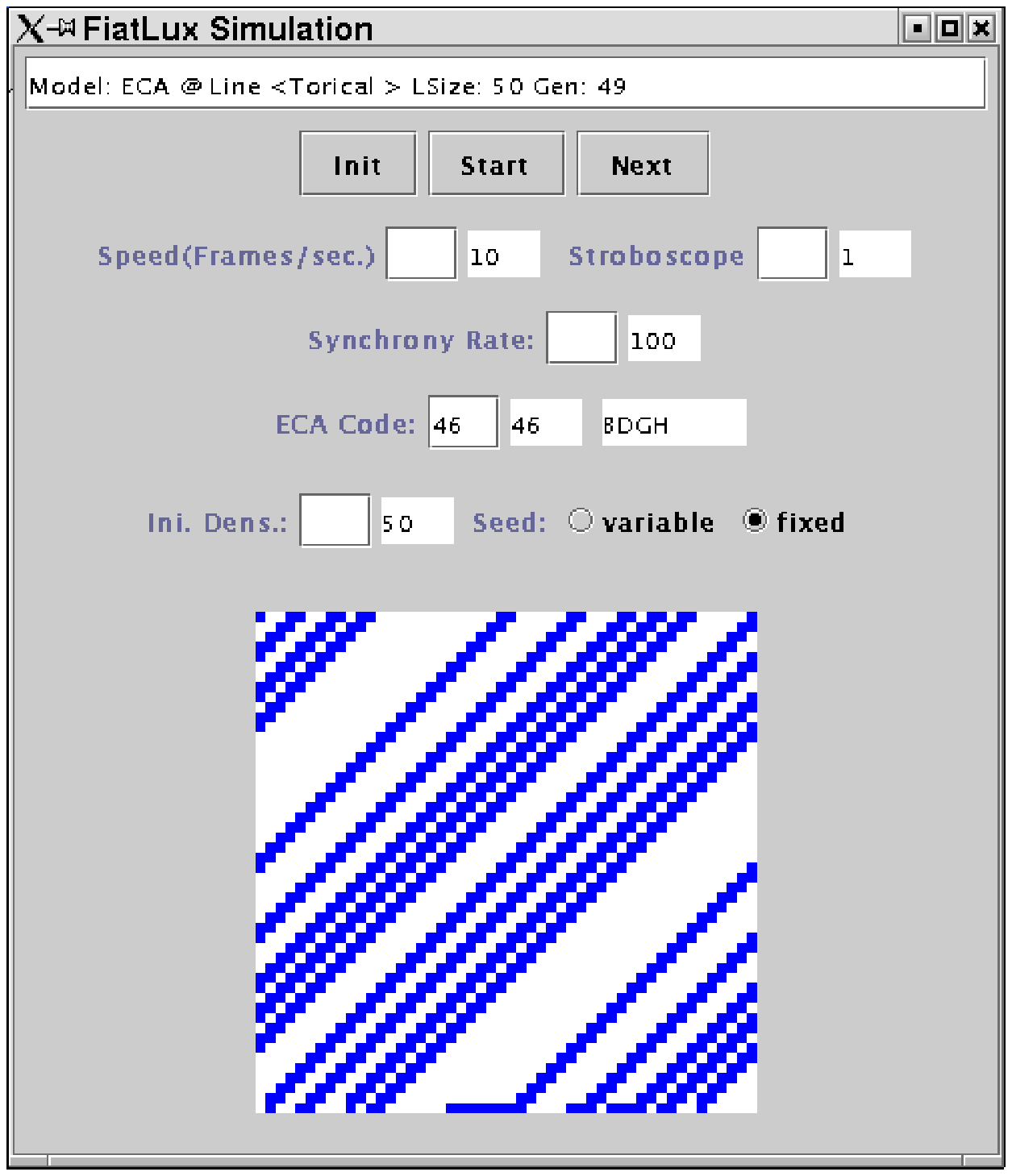}{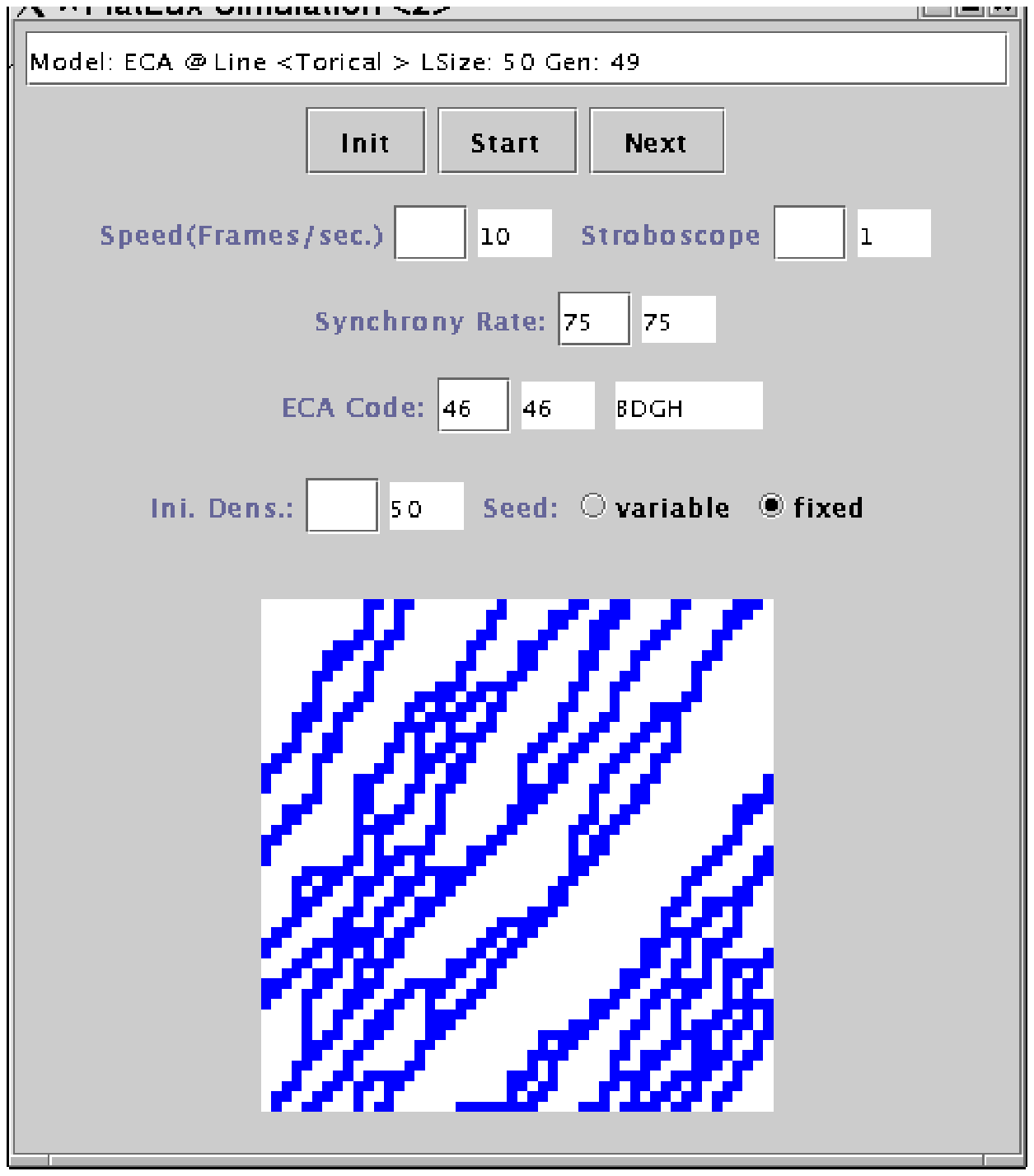}{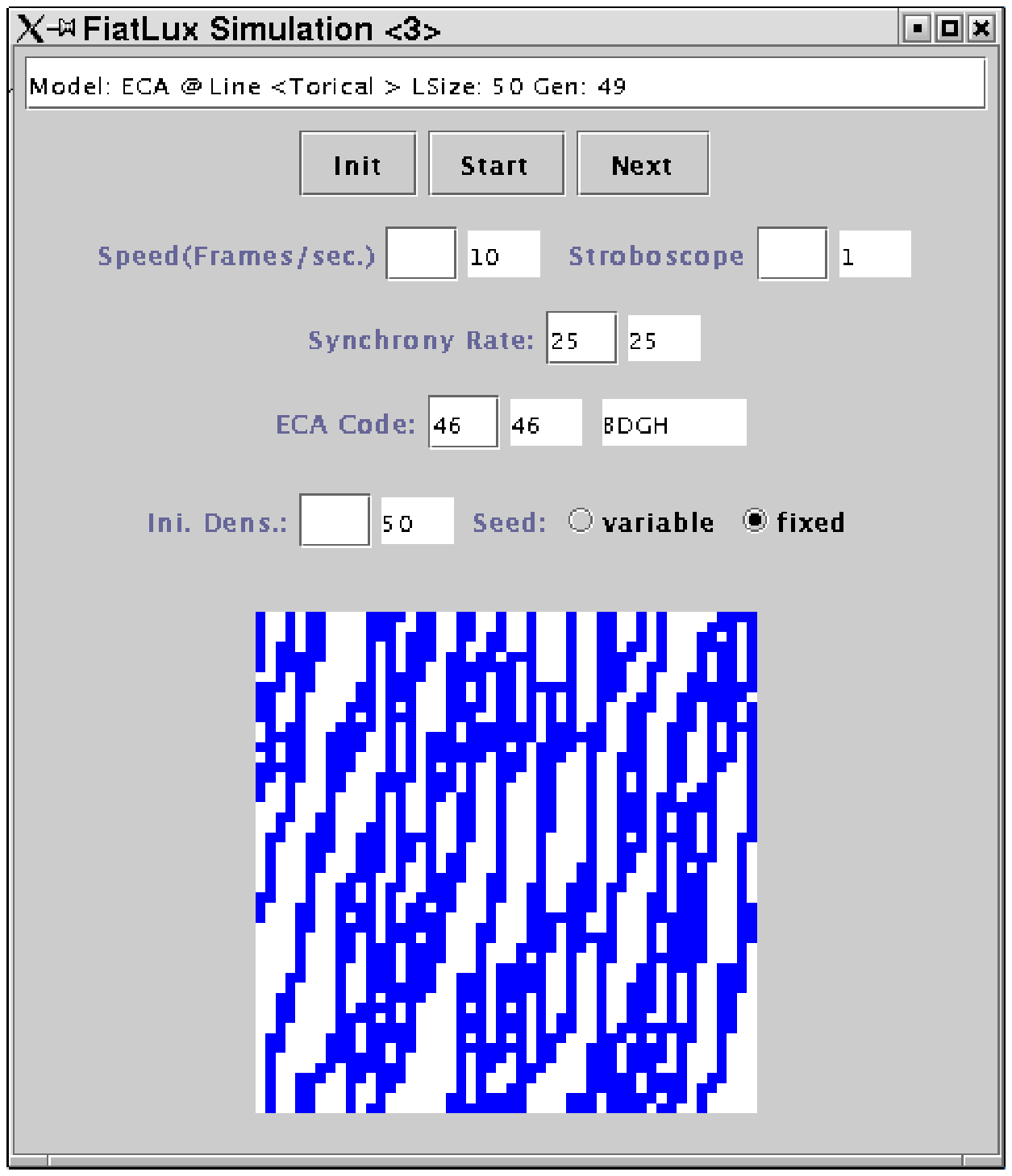}
\DoubleCaptionLabel{\LegSurfRiddlesa}{\LegSurfRiddlesb}{ECA46lbl}
\EF


\begin{figure}[htbp]
\centering
\LargeOrbit{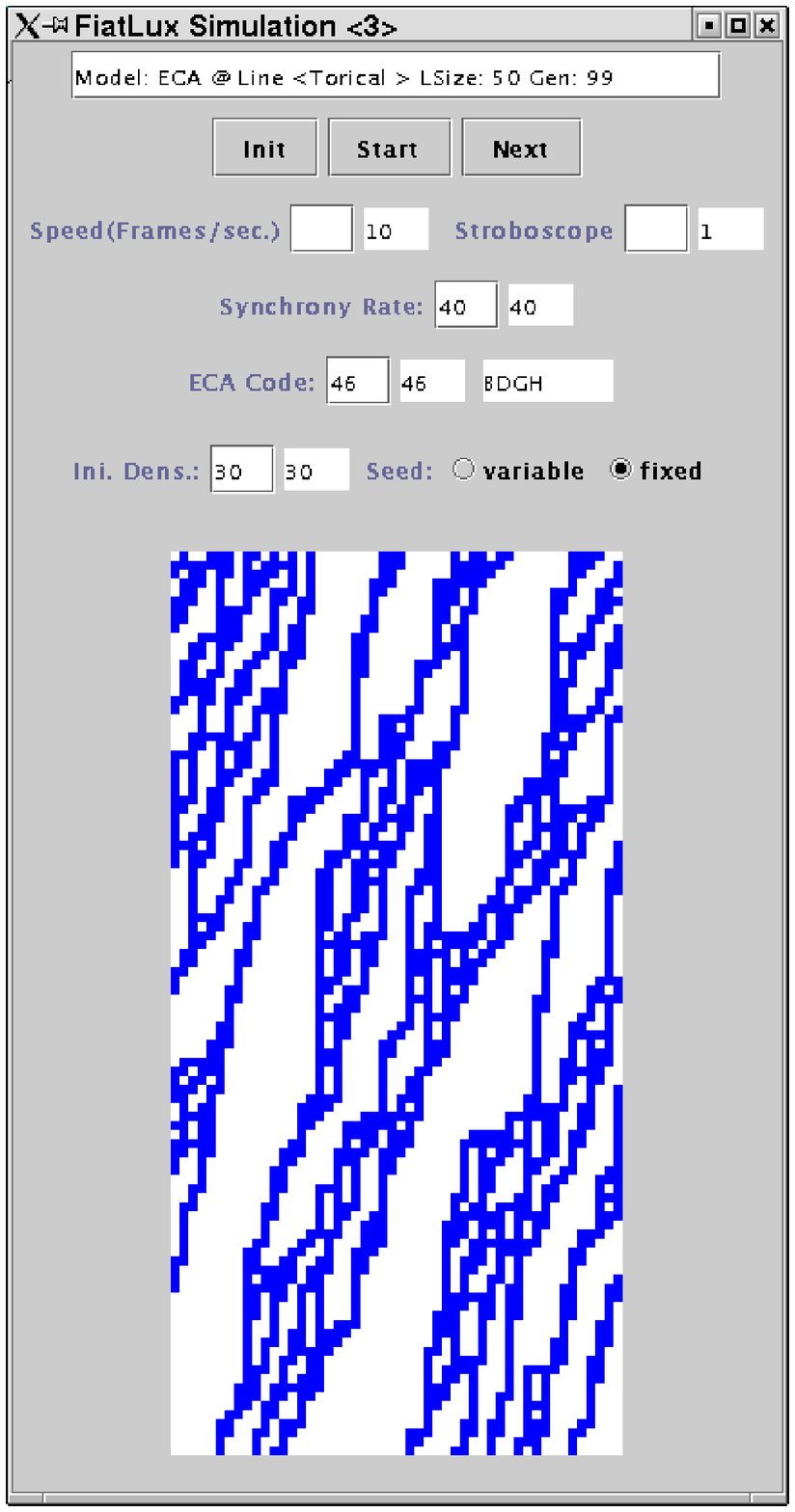}
\LargeOrbit{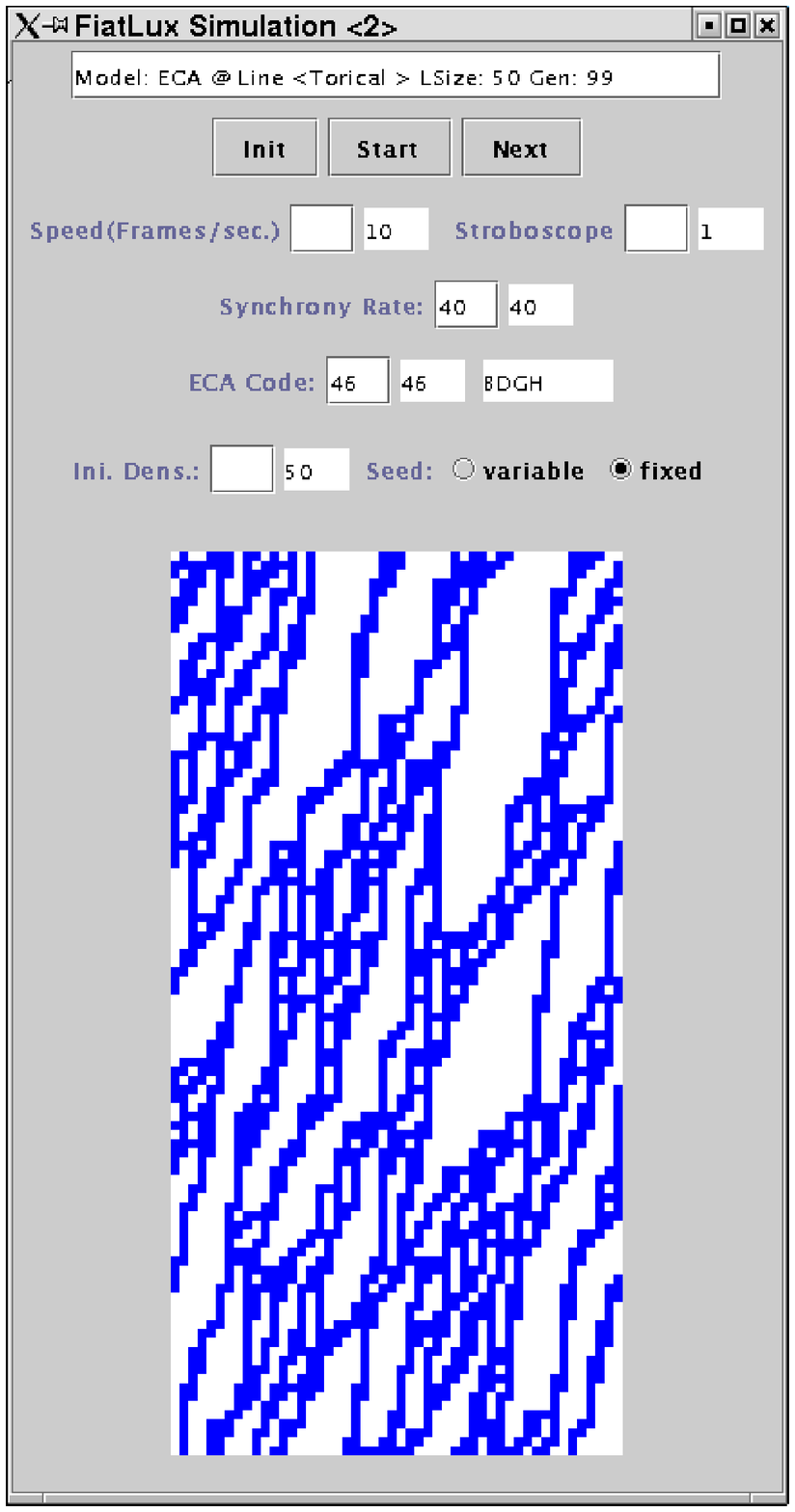}
\LargeOrbit{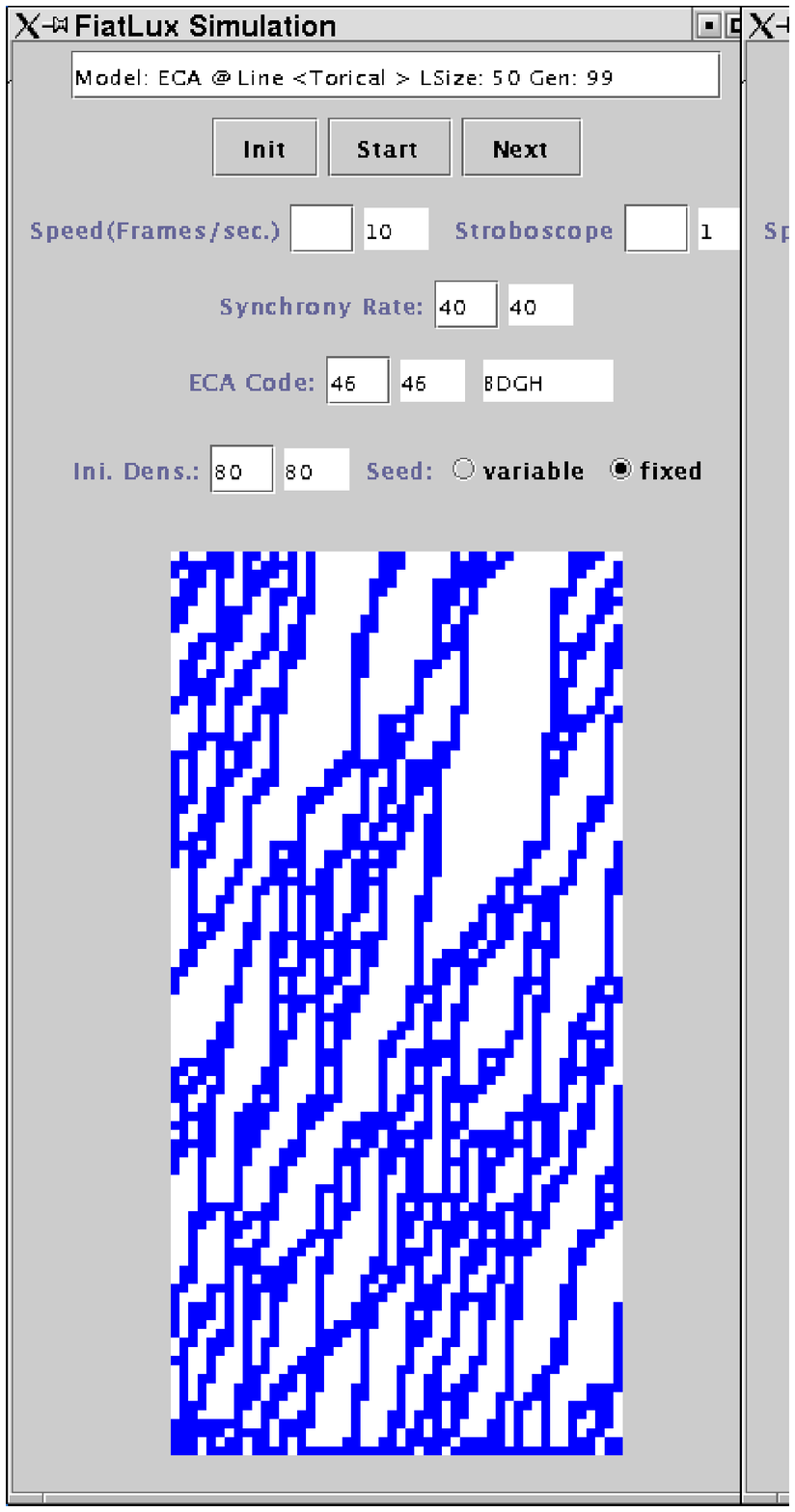}
\caption{Evolution of ECA \ECA{46} for $ \sr= 0.40 $ and $ \dini $ = 0.30 (left) 0.50 (center) 0.80 (right) }
\label{Fi:ECA46orbD}
\end{figure}

The examination of the sampling surface for \ECA{46} revealed a surprising phenomenon~: 
``riddles'' almost parallel to the $\dini$-axis appear on the sampling surface (see \figref{ECA46lbl}).
We conjecture that ECA \ECA{46} is a model for which there exists a subset of configurations $ I \subset \E $ which provide ``merging orbits''~:
\[ \forall (x_1,x_2) \in F \times F, \forall \sr \in ]0,1], \exists t, \gamma_{\sr}(x_1,t) = \gamma_{\sr}(x_2,t) = x_t \COMMA \]
with the particularity that $ x_t $ is not a fixed point. 
This can be observed in \figref{ECA46orbD} in which $ \sr $ is kept constant and where $ \dini $ varies.

The very existence of such models is surprising since it implies that different initial conditions eventually merge into the same 
orbit without even stabilizing on a fixed point.
Obviously for ECA \ECA{46}, $ I $ is not strictly equal to $ \E $ as $ \zero $ is not part of $ I $ (it is a fixed point).
However, informal experiments starting from various initial conditions lead to conjecture that $ I = \E - \set{\zero} $ 
meaning that for a fixed dynamics, all non-zero configuration eventually merge into a single orbit.
Such result should be explored in a future work both by experimental and formal approach.

\subsubsection{$ \dini $-invariant,``U''-shaped surfaces}

\BF
\SSurf{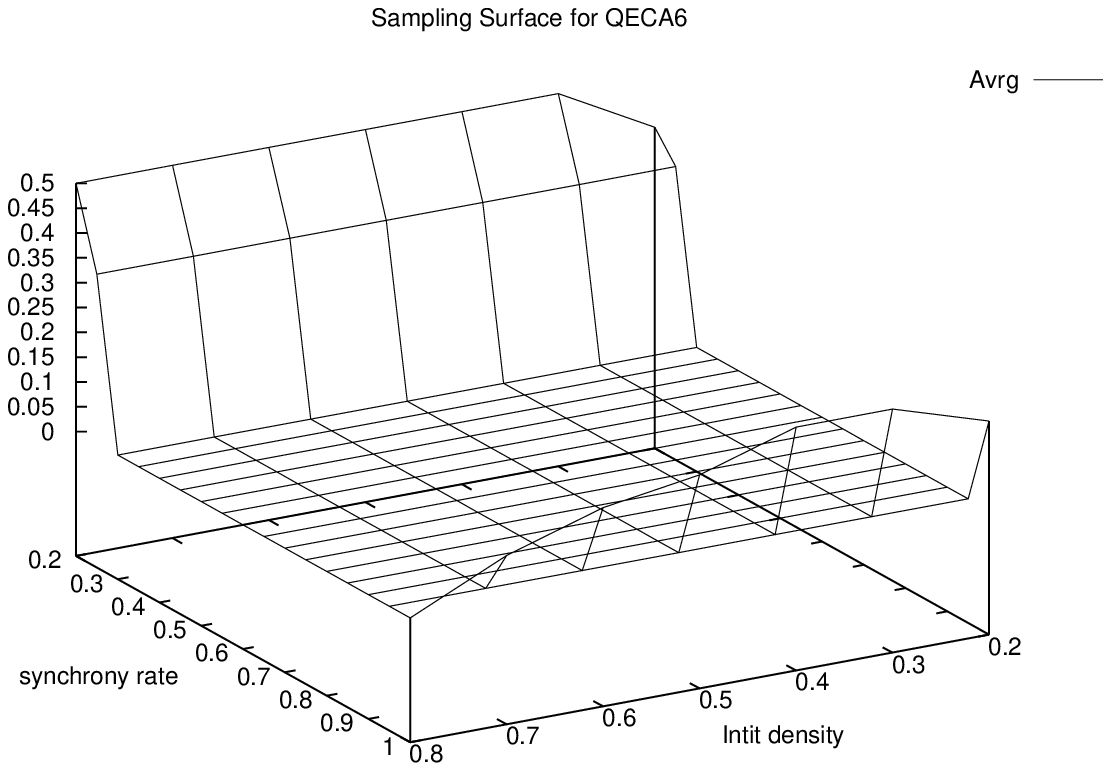}
\TripleOrbit{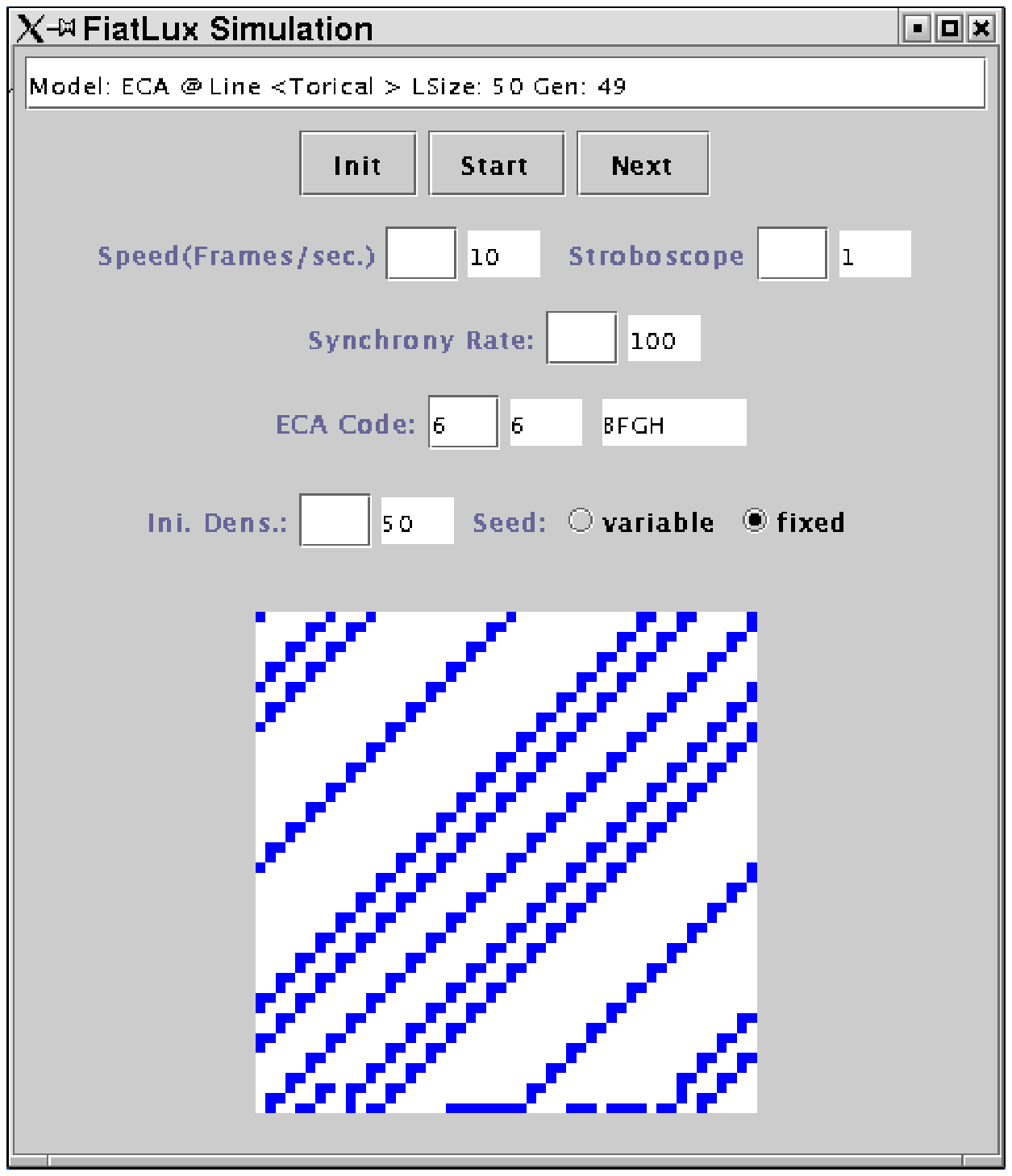}{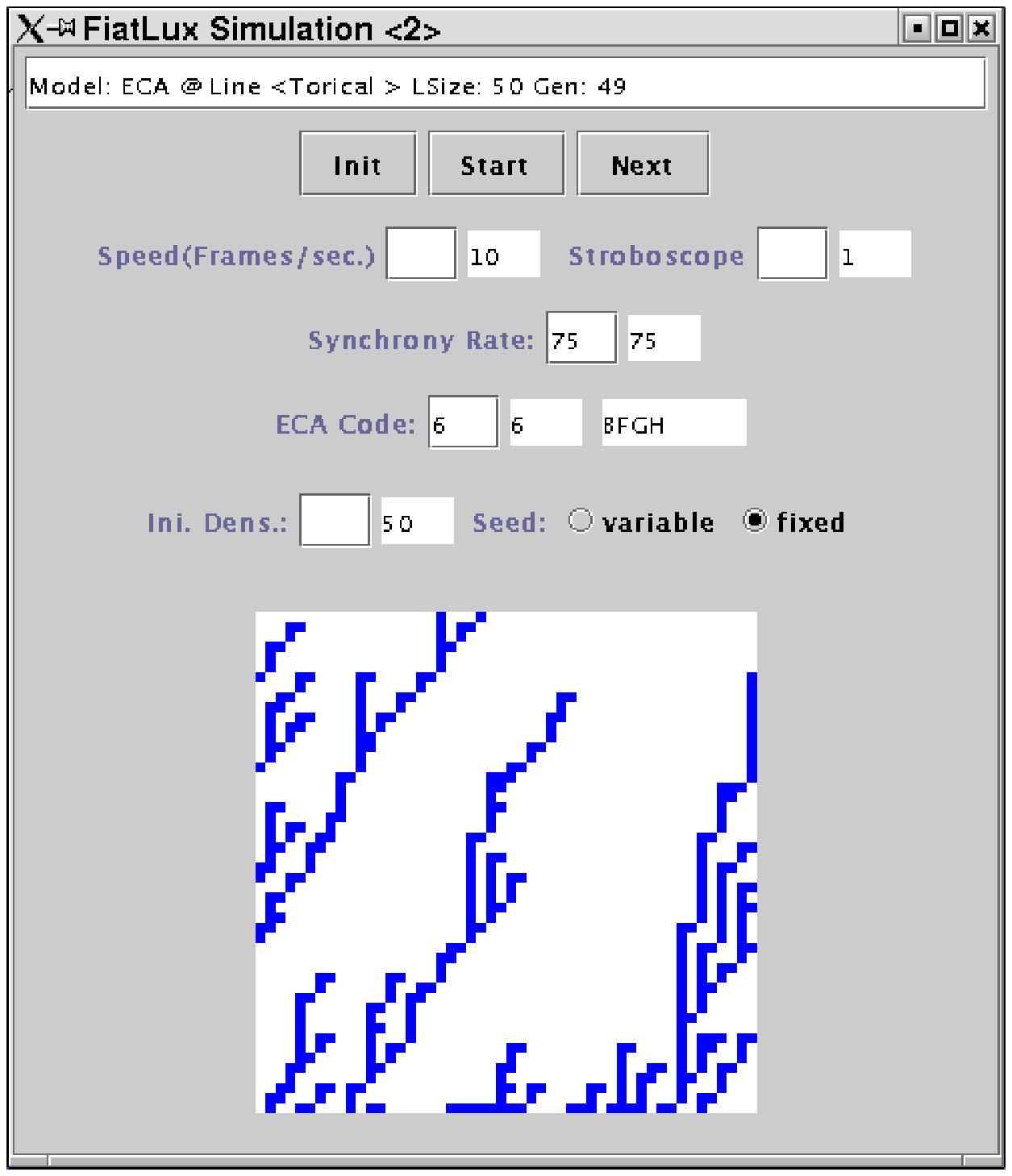}{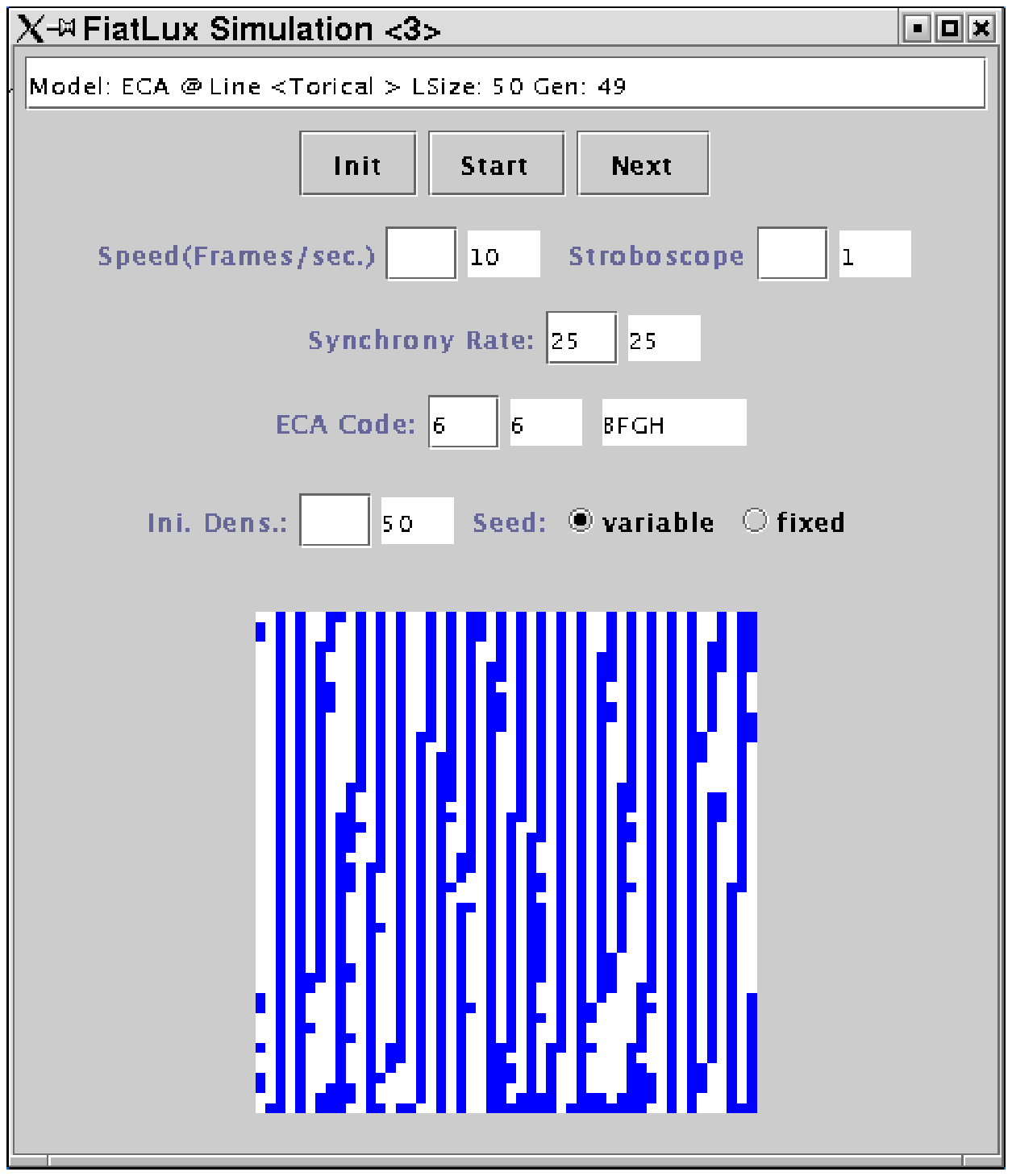}
\DoubleCaptionLabel{\LegSurfUa}{\LegSurfUb}{ECA6lbl}
\EF

ECA \ECA{6}, \ECA{38} and \ECA{134} have an unexpected behavior~: 
just like GAP the introduction of a little bit of asynchronism makes the system evolve to a homogeneous fixed point. 
However, unlike SPT ECA, the observation of a long-lived branching structure occurs for values of $ \sr $ {\em smaller} than $ \sr_c $. 

ECA \ECA{6} sampling surfaces illustrates how GAP-type discontinuity at $ \sr = 1 $ and 
an SPT-type discontinuity at $ \sr \eq 0.3 $ (see \figref{ECA6lbl}) can both cohabitate.
The conjunction of both characteristics explains why this model is situated in zone D (high $ \ra $, high $ \rb $).
It is worth noticing that the unstable phase ($ \muf > 0 $) is obtained for values of synchrony rates that are {\em lower} than 
the critical value $ \sr_c$ and the stable phase (fixed point $ \zero$, $ \muf = 0$) is located for $ \sr > \sr_c $.
It implies that the system can become {\em less} stable when asynchronism is increased.
This observation seems to contradict the thesis proposed in \cite{Ber94} which conjectured that the increase of asynchrony
has a stabilizing effect on the dynamics of the models. \label{Sec:Bersini} 
It shows that a deeper analysis is needed to understand when the increase of asynchrony (i.e., the decrease of $ \sr $) 
may stabilize a model by allowing it to reach a fixed point or a stable phase. 

\section{ Discussion } \label{Sec:Discussion}

In this paper we described a general-purpose scheme to quantify the robustness of a CA to asynchronism. 
We chose to observe this robustness according to a protocol which used the density macroscopic parameter and 
a sampling strategy based on choosing randomly initial conditions and synchrony rates.
We have applied this protocol to the 88 equivalence classes of the ECA space 
to show that a wide variety of phenomena could be observed.
In order to go further than the simple visual observation of the orbits we used the sampling surfaces as a synthetic means of representing a model's robustness and we proposed two indicators to induce a partial order on the models by quantifying this robustness in $ \R^2 $. 
This methodology allowed us to induce a distinction between the different rules of the ECA space and 
to define robustness classes according to the types of changes that were observed when we added asynchronism in the update rule.
We can now discuss our initial questions in two directions : about robustness and about modeling.

\subsection{About robustness}

An important feature of our classification is that the classes defined according to robustness criteria
cannot be deduced from Wolfram's empirical classification \cite{Wol84}.
For example, if we take the ``chaotic'' rules, we find that ECA \ECA{122} is in Zone A while \ECA{18} and \ECA{146} are in Zone C (SPT).
If we take the ``periodic'' rules, we find that ECA \ECA{232} is in Zone A, \ECA{34} is in Zone B (GAP), 
\ECA{50} is in Zone C (SPT), rule \ECA{6} is in Zone D (U-shaped).
This opens new perspectives for constructing a theory 
which could predict the shape of the sampling surfaces by analyzing the form of the local transition rule.
We proposed the use of walls as a first step in this analysis with the ECA \ECA{232} and \ECA{73}.
This classification based on robustness might equally be related with the classification proposed by K\accent23 urka\cite{kurka97}.
Indeed, it has been shown that the existence of blocking words allows one to determine the class of an automaton and 
it appears that walls are just a stronger version of blocking words. 
It has been recently demonstrated that at least three of the four classes of this classification are undecidable \cite{Var03} 
but the question remains open to decide whether a classification based on walls might be decidable and easily computable. 

It is important to notice that we never used the fact that the analyzed objects were two-state, radius one, one-dimensional CA 
in the definition of the experimental protocol. 
This leaves the possibility to explore the behavior of models defined with a higher number of states and in higher dimensions.
For two dimensional CA, the study of robustness could be as well examined with respect to changes in the lattice topology. 
Indeed, one may also want to know whether a small perturbation on the regularity of the lattice 
may produce significant changes in the behavior of a 2D cellular automaton.

Another possibility of improving the study concerns a finer evaluation of the quality of the statistics. 
In our protocol, the number of initial conditions chosen for the sampling is relatively small ($ \leq 100 $)
and do not allow us to detect interesting particular subsets of configurations which may produce different results.
This suggests that once a model is declared robust (Zone A), 
it should be studied for a large number of initial conditions 
to quantify precisely the fraction of initial conditions for which robustness is observed. 
This could be done with analytic methods or with an exhaustive experimental study of small ($ \D \leq 30 $) ring sizes 
and would provide further refinements of the classification.


\subsection{About complex systems modeling}

The experimental method developed here is a first approach that can be used as a guideline to select suitable rules
for complex systems modeling with cellular automata.
The results presented in this work showed that according to the wide range of phenomena observed when asynchrony is introduced,
the use of CA as a modeling tool could take advantage of the classification into robustness zones :  
\BL
\item The analysis of Zone A allowed us to find the rules which could be suitable for modeling~: 
they show stability to the perturbation to asynchronism according to some observation function. 
A strong version of robustness was found in the PR models which obeyed a stronger robustness criterium : for all the initial condition tested, 
the same asymptotic behavior was reached whatever the value of the synchrony rate.
The use of such models may provide a way of building CA-based devices with a behavior strongly tolerant to asynchronism.

\item The behavior of Zone B models, and particularly the GAP class, 
suggests that their synchronous should be discarded for a real-world application, 
except if the purpose of the model is precisely to detect the existence of asynchronism. 
However, in the asynchronous regime, they appear very stable as the same asymptotics are reached whatever the initial conditions.

\item Identically, some Zone C rules showed that a brutal change in their behavior could occur 
for a particular critical value $ \sr_c $ of asynchronism. 
This kind of effect can be undesirable if the modeled phenomenon is not supposed to be synchrony-dependent. 
On the other hand, one may want such feature to be exhibited by a model. 
For example, in biology, it is known that the aggregation of the {\em Dictyostelium Discoidum} 
is triggered when a critical value of starvation is reached. 
To our knowledge, none of the various models (e.g., \cite{Sav97} )proposed yet have been successful 
in predicting the existence of such a critical value. 
The explanation could be that the release of a chemical component (cAMP) changes the ``synchronicity'' between cells 
and that the communication between cells is directed by percolation-like effects that explains why the triggering of the aggregation is sudden. 
In social sciences, a model used for understanding urban settlement also showed great disparities between the synchronous and asynchronous behavior, 
the ``synchrony rate'' here being controlled by the ``mobility'' (ability to go and live elsewhere) of the agents \cite{Van00}.

\item The existence of models in Zone D indicate that despite the spatial and temporal averaging we used in the definition 
of the observation function that quantifies a CA behavior, the outcome of the experiments remained irregular. 
Such models show that the behavior of an ACA may be simple when evolved synchronously and much more complex with an asynchronous update rule (e.g, the shift).

\EL

The phenomenology we observed and the existence of robust CA rules suggests that we can no longer claim that a CA model is not valid because transitions occur too regularly to capture real-world phenomena~: 
even though the ``real-world cells'' might affected by some permanent irregularities (synchronism and/or topology faults) or by noise, 
a CA model might be robust enough to produce the same output when evolved with perturbations.
This further suggests that there exists no universal answer to the question 
of knowing which part of the interesting behavior of a (classical) CA is due to the synchronism. 
Each modeling problem should instead be studied with a specific approach and the macroscopic parameters and observation functions used in this work, 
far from being universal, should be chosen according to what feature of the CA is desired to be robust. 
For example, one may interested in using a CA with many states to model propagating signals in an excitable medium. 
In this case, one should find the suitable parameters to assess the ability to propagate signals and use these parameters in the robustness assessment.

\section*{Acknowledgments}

We wish to thank Mats Nordahl (University of G\"oteborg, Sweden) for the stimulating discussions held during the Exystence Thematic Institute, 
Cristopher Moore (Santa Fe Institute, USA), Marianne Delorme, Jacques Mazoyer, Bertrand Nouvel and Fr\'ed\'eric Chavanon (ENS Lyon, France) 
for their advice and reading.
The LIP is the parallelism computer science laboratory of ENS Lyon; 
it is associated with the CNRS, the ENS Lyon, the INRIA and the University Claude Bernard Lyon I.

{\small
\bibliographystyle{amsplain}
\bibliography{ArticleRobustesse.bib}
}

\end{document}